\documentclass[journal]{IEEEtran}
\usepackage[british]{babel}
\usepackage{cite}
\usepackage{amsmath,amssymb,amsfonts,amsthm}
\usepackage{algorithmic}
\usepackage{physics}
 \usepackage{graphicx}
\usepackage{textcomp}
\usepackage[T1]{fontenc}
\usepackage{caption}
\usepackage{cases}
\usepackage{subcaption}
\usepackage{epstopdf}
\usepackage{overpic}
\usepackage{array}
\usepackage{color}
\usepackage{url}
\usepackage{bbm}
\usepackage{enumerate}
\usepackage[shortlabels]{enumitem}

\usepackage[none]{hyphenat}
\usepackage{hyperref}
\usepackage{lipsum}

\def\BibTeX{{\rm B\kern-.05em{\sc i\kern-.025em b}\kern-.08em
    T\kern-.1667em\lower.7ex\hbox{E}\kern-.125emX}}
    
\usepackage{color}
\usepackage[dvipsnames,svgnames]{xcolor}
\usepackage[textwidth=30mm]{todonotes}

\def\Htran{\mbox{\tiny $\mathrm{H}$}}
\def\Ttran{\mbox{\tiny $\mathrm{T}$}}

\def\Real{\mathbb{R}}
\def\Complex{\mathbb{C}}

\def\imagunit{\mathsf{j}} 

\newcommand{\vect}[1]{{\boldsymbol{#1}}}

\DeclareMathOperator*{\dprime}{\prime \prime}

\theoremstyle{plain}

\usepackage[margin=0.93in]{geometry}

\IEEEoverridecommandlockouts

\begin{document}

\title{\Huge{Wide-Aperture MIMO via \\Reflection off a Smooth Surface}}

\author{\vspace{-0.0cm}
\IEEEauthorblockN{Andrea Pizzo, \emph{Member, IEEE}, Angel Lozano, \emph{Fellow, IEEE}, \\Sundeep Rangan, \emph{Fellow, IEEE}, Thomas L. Marzetta, \emph{Life Fellow, IEEE}\vspace{-.0cm}
\thanks{\vspace{0.0cm}
\newline \indent 
Parts of this work were presented at the IEEE Veh. Techn. Conf. (VTC'22 Spring) \cite{PizzoVTC22}.
A.~Pizzo and A.~Lozano are with Univ. Pompeu Fabra, 08005 Barcelona, Spain (email: \{andrea.pizzo, angel.lozano\}@upf.edu). 
S.~Rangan and T.~Marzetta are with New York University, 11201 New York, USA (email: \{s.rangan, tom.marzetta\}@nyu.edu).
Work supported by the European Research Council under the H2020 Framework Programme/ERC grant agreement 694974, by the ICREA Academia program, by the European Union-NextGenerationEU, and by the Fractus-UPF Chair on Tech Transfer and 6G.
}
}}


\maketitle
\vspace{0.0cm}
\begin{abstract}
This paper provides a deterministic channel model for a scenario where wireless connectivity is established through a reflection off a smooth planar surface of an infinite extent. The developed model is rigorously built upon the physics of wave propagation and is as precise as tight are the unboundedness and smoothness assumptions on the surface. This model allows establishing how line-of-sight multiantenna communication is altered by a reflection off an electrically large surface, a situation of high interest for mmWave and terahertz frequencies.
\end{abstract}


\section{Introduction}

The wealth of unexplored spectrum in the millimeter wave (mmWave) and terahertz 
 ranges brings an onrush of wireless research seeking its fortune at higher frequencies \cite{Sundeep2014,Rappaport2015,Rappaport2019}.
The short range for which these frequencies are most suitable, in conjunction with the tiny wavelength, enable reasonably sized arrays to access multiple spatial degrees of freedom (DOF) even in line-of-sight (LOS) \cite{Rodwell2011}. 
Precisely, LOS spatial multiplexing is made possible by the rich pattern of phase variations of the radiated field's spherical wavefront, which mimics the diversity richness of multipath propagation at lower frequencies.
This potential has unleashed much research activity on wide-aperture multiple-input multiple-output (MIMO) communication over LOS channels \cite{Bohagen,9422343}.

A downside of these high frequencies is blockage and lack of diffraction around obstacles, which may render LOS MIMO vulnerable to interruptions.
This naturally raises the interest in studying whether wide-aperture MIMO could also operate through a reflection, capitalizing on the availability in many environments of interest of surfaces that are electrically (i.e., relative to the wavelength) large.

This paper seeks to examine MIMO communication via reflection off a smooth planar surface of infinite extent. To this end, one possibility would be to apply ray-tracing tools \cite{Rangan2022}, but the accuracy to which the environment should be characterized to prevent artifacts is not known a priori. Also, ray tracing does not provide analytical insights into the underlying propagation mechanisms, which are essential to array optimization.
Instead, we derive a deterministic physics-based scalar channel model that is valid irrespective of the communication range and embodies other models as particular cases.

\subsection{Contributions}

Although an actual reflecting surface is necessarily finite and with some degree of roughness, at sufficiently high frequencies it may be reasonably regarded as infinitely large, as the impact of diffraction 
vanishes. Oppositely, the roughness is emphasized at high frequencies as irregularities on the surface become comparable to the small wavelengths. The latter aspect is not considered in this paper, left for future work. 

Motivated by the extensive physics literature on the interaction between a plane wave and an infinite smooth surface \cite{OpticsBook,ChewBook}, we start by expanding the 3D field generated by an arbitrary source in terms of plane waves
\cite{PlaneWaveBook,ChewBook}. Fundamental principles describing the reflection and transmission phenomena at the surface can then be applied to each plane wave separately and combined to obtain the overall field at any point \cite{ChewBook}. 
%
%
An LOS channel is seen to be the cascade of a low-pass filter that cuts off evanescent waves \cite{PizzoTSP21}, and a reverse-bowl-shaped filter imposed by the wave equation \cite{PizzoJSAC20}; a reflection off a surface adds an additional filtering stage that augments the model in \cite{PizzoTSP21,PizzoJSAC20} with backward propagation.
This paper can also be seen to complement the zero-mean stochastic model derived in \cite{PizzoIT21}, with their conjunction yielding a Rician fading model.

After discretization through spatial sampling, a deterministic description of the channel is obtained. This is finally used to numerically evaluate the eigenvalues, DOF, and spectral efficiency for the purpose of MIMO communication. 
Altogether, the contributions are:
\begin{itemize}
\item
Starting from first principles, a channel model is developed that builds upon the physics of wave propagation. The analysis is as precise as tight are the unboundedness and smoothness assumptions on the surface.
\item
Progress is made, in the wake of \cite{PizzoJSAC20,PizzoIT21,PizzoTSP21}, towards a comprehensive physics-based modeling of wireless propagation on which signal processing and communication theorists can test their algorithms. Propagation is described in terms of spatial Fourier transforms and linear system theory, notions central to both communities.
\item
Classical electromagnetic results such as the image theorem 
are revisited. These have fundamental implications on the optimization of antenna spacings as a function of the signal-to-noise ratio (SNR) and they allow extending results available for a pure LOS channel \cite{HeedongISIT,Heedong2021} to a reflection channel.
\end{itemize}


\subsection{Outline and Notation}

The manuscript is organized as follows. Sec.~\ref{sec:propagation} revisits the physics behind plane-wave reflection off a smooth planar surface 
relying solely on linear system theory and Fourier transform.
In Sec.~\ref{sec:Fourier_pw}, the Fourier spectral representations of the LOS and reflected  transmissions are derived. The connection with the image theorem is established in Sec.~\ref{sec:image_th}, whereas the channel impulse response follows in Sec.~\ref{sec:channel_response}.
After discretization, the channel response is used in Sec.~\ref{sec:numerical_results} to assess the MIMO performance via reflection. A comparison with ray-tracing is presented in Sec.~\ref{sec:ray_tracing}.
Final discussions and possible extensions are set forth in Sec.~\ref{sec:conclusions}. 
   
We use upper (lower) case letters for spatial-frequency (spatial) entities
while 
$J_0(\cdot)$ is the Bessel function of the first kind with order $0$, $(x)^+ = \max(x,0)$, and $\delta(\cdot)$ is the Dirac delta function. 

\section{Plane-Wave Interaction with Materials} \label{sec:propagation}
       
Narrowband propagation is considered at angular frequency $\omega$ in a 3D medium with an inhomogeneity created by a $z$-oriented planar object of infinite thickness,
dividing the medium into a region~1 ${\{r_z<0\}}$ (free space) and a region~2 ${\{r_z>0\}}$ (material). 
The electromagnetic properties are constant in each of the two ensuing regions, characterized by the refractive indexes $n_1=1$ and $n_2 \in \Complex$ with $\Re(n_2)\ge 1$ and $\Im(n_2)>0$ modeling the phase variations and absorption losses occurring inside the material \cite[Sec.~4.2]{MolischBook}.
The wavenumbers in the two regions are $\kappa_1 = 2\pi/\lambda$ and
\begin{equation} \label{wavenumbers}
 \kappa_2 = n_2 \kappa_1.
\end{equation}

\subsection{Dielectric Half-Space}

 \begin{figure}
\centering\vspace{-0.0cm}
\includegraphics[width=.999\linewidth]{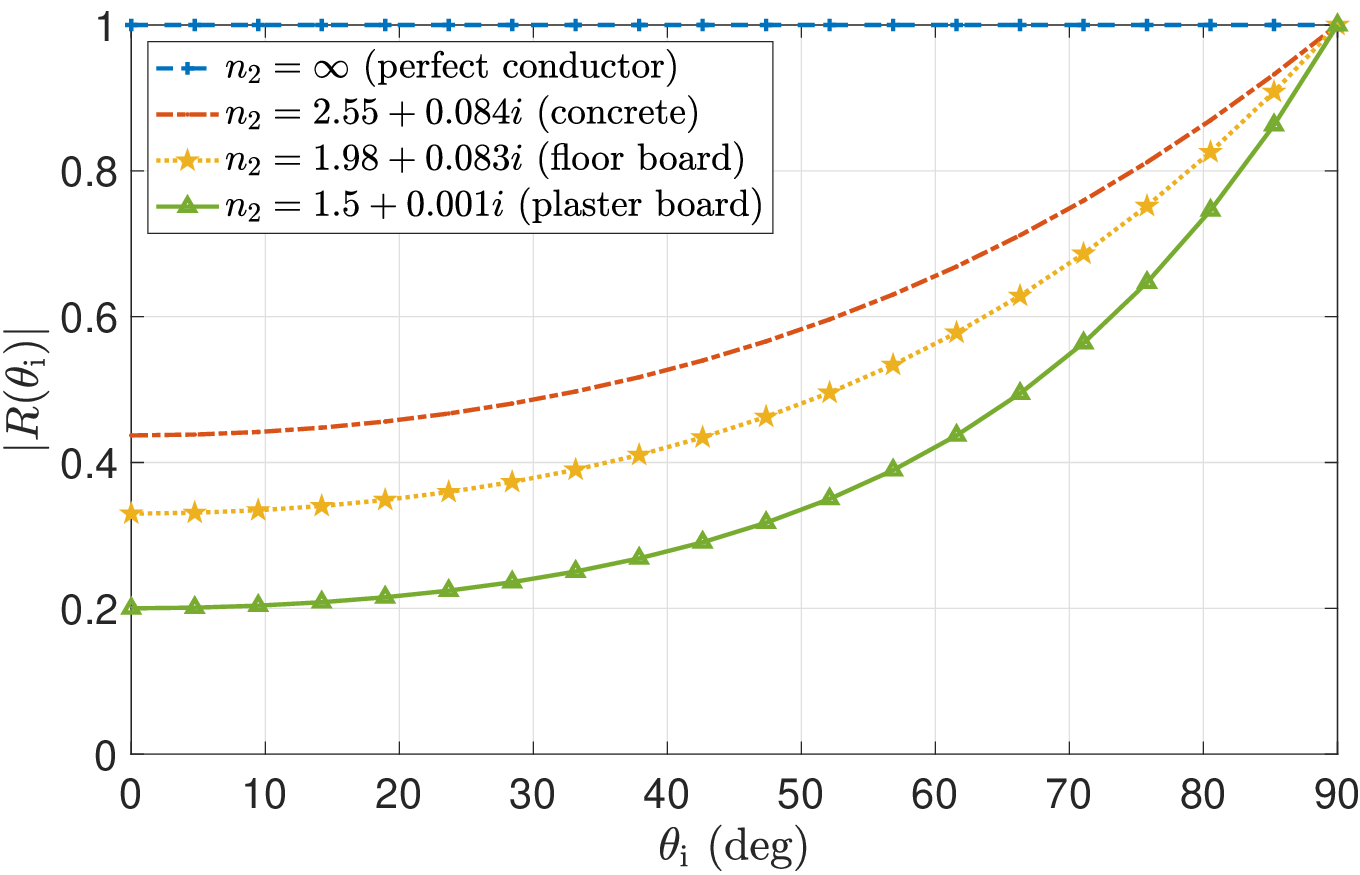} 
\caption{Fresnel reflection coefficient (magnitude) as a function of ${\theta_{\text i}}$ for various refractive indices.}\vspace{-0cm}
\label{fig:Fig1}
\end{figure}
        
We first consider the $xz$-plane containing the direction of propagation and the surface normal, namely the \emph{plane of incidence}.\footnote{This plane can always be obtained by rotating the Cartesian reference frame opportunely about the $x$-axis.}
A point in this plane has coordinates $(r_x,r_z)$. 
An upgoing \emph{incident} plane wave
\begin{equation} \label{incident_wave}
e_{\text i}(r_x,r_z) =  E_{\text i}(\theta_{\text i}) \, e^{\imagunit \kappa_1 \left(r_x \sin\theta_{\text i} +  r_z \cos\theta_{\text i} \right)}
\end{equation}
with amplitude $E_{\text i}(\theta_{\text i})$ traveling in region~1 from an angle $\theta_{\text i}$ relative to the surface normal impinges thereon.
As a result of interaction with the surface, this field creates a downgoing \emph{reflected} plane wave in region~1,
\begin{equation} \label{reflected_wave}
e_{\text r}(r_x,r_z) = E_{\text r}(\theta_{\text r}) \, e^{\imagunit \kappa_1 \left(r_x \sin\theta_{\text r} - r_z \cos\theta_{\text r} \right)},
\end{equation}
with amplitude $E_{\text r}(\theta_{\text r})$ and angle $\theta_{\text r}$ and another upgoing \emph{transmitted} plane wave in region~2,
\begin{equation} \label{transmitted_wave} 
e_{\text t}(r_x,r_z) = E_{\text t}(\theta_{\text t}) \, e^{\imagunit \kappa_2 \left(r_x \sin\theta_{\text t} + r_z \cos\theta_{\text t} \right)},
\end{equation}
with amplitude $E_{\text t}(\theta_{\text t})$ and angle $\theta_{\text t}$. 
Derivable from the boundary conditions, Snell's law dictates that reflection occurs at the specular angle $\theta_{\text r} = \theta_{\text i}$ while transmission is specified by $\sin(\theta_{\text t}) = \sin(\theta_{\text i})/n_2$ \cite[Eq.~1.5.6]{OpticsBook}.
The complex-valued plane-wave amplitudes can be written in terms of the \emph{Fresnel coefficients} $R(\theta_{\text i}) = E_{\text r}/E_{\text i}$ and $T(\theta_{\text i}) = E_{\text t}/E_{\text i}$, specifying the fraction of incident field reflected from or transmitted across the surface, for every incident angle. Their magnitude is always less than unity, and they satisfy the unitarity relation $T(\theta_{\text i}) = 1 + R(\theta_{\text i})$ 
due to conservation of energy.

Multiple reflections that might arise inside an object of finite thickness would make the interaction with the surface more involved \cite[Ch.~2.1.3]{ChewBook}. However, these never occur at  frequencies high enough such that the material thickness is much larger than the wavelength, making the reflection phenomenon highly predictable and suitable for array optimization, as will be seen.

The complex-valued Fresnel reflection coefficient is given by \cite[Eq.~7.4.2]{OrfanidisBook}\footnote{For every angle $\theta_{\text i}$ there are two linearly independent plane waves being the solutions of the two scalar wave equations characterizing the transverse electric (TE) polarization, where the electric field is parallel to the surface, and the transverse magnetic (TM) polarization, where the magnetic field is parallel \cite[Ch.~2.1]{ChewBook}. We concentrate on the TE equation as the TM’s is obtainable by invoking the duality principle.}
\begin{equation} \label{R_theta}
R(\theta_{\text i}) = \frac{ \cos(\theta_{\text i}) -  \sqrt{n_2^2 - \sin^2(\theta_{\text i})}}{ \cos(\theta_{\text i}) +   \sqrt{n_2^2 - \sin^2(\theta_{\text i})}},
\end{equation}
whose magnitude is plotted in Fig.~\ref{fig:Fig1} as a function of ${\theta_{\text i}}$ for various dielectric materials \cite{Sato}. 
Total reflection is achieved by a perfect conductor, 
which behaves as a mirror. 
Other materials behave as perfect conductors only at a grazing incidence. 
In general, denser materials reflect energy better and, for a given material, close-to-grazing incidences experience higher reflections than those near the normal.

\subsection{Linear-System-Theoretic Interpretation}

We now deviate from physics and provide a different viewpoint on the interaction mechanism with the surface; this perspective relies only on linear system theory and Fourier transforms, key results in the toolbox of communication theorists.

The propagation directions of the incident, reflected, and transmitted plane waves may alternatively be specified by the wavenumber coordinates
\begin{align} \label{wavenumber_coord}
(\kappa_x, \pm \kappa_{1z}) & = (\kappa_1 \sin\theta_{\text i},\pm \kappa_1 \cos\theta_{\text i})  \\
(\kappa_x, \kappa_{2z}) & = (\kappa_2 \sin\theta_{\text t}, \kappa_2 \cos\theta_{\text t}) 
\end{align}
satisfying the dispersion relations $\kappa_x^2 + \kappa_{iz}^2 = \kappa_i^2$ for $i=1,2$. 
By means of \eqref{wavenumber_coord}, the plane waves in \eqref{incident_wave} and \eqref{reflected_wave} can be seen as the 2D Fourier harmonics
\begin{align}
e_{\text i}(r_x,r_z) & =  E_{\text i}(\kappa_x) \, e^{\imagunit \left( \kappa_x x +  \kappa_{1z} z \right)} \\
e_{\text r}(r_x,r_z) & =  E_{\text r}(\kappa_x) \, e^{\imagunit \left( \kappa_x x -  \kappa_{1z} z \right)},
\end{align}
which are functions of the spatial-frequency variables $(\kappa_x,\kappa_{1z})$. The same holds for \eqref{transmitted_wave}, expressed as
\begin{equation}
e_{\text t}(r_x,r_z) =  E_{\text t}(\kappa_x) \, e^{\imagunit \left( \kappa_x x + \kappa_{2z} z \right)}
\end{equation}
for  $(\kappa_x,\kappa_{2z})$. 
The connection with Fourier theory that the above change of variables establishes enables a linear-system-theoretic interpretation of the reflection and transmission phenomena, with the focus henceforth being on the reflection.

The response to a harmonic input at spatial frequency $(\kappa_x,\kappa_{1z})$ is another harmonic output at the same spatial frequency---up to a change of sign in $\kappa_{1z}$ due to the reflected wave traveling in the opposite direction---whose
complex amplitude is the product of the input's amplitude and the Fresnel spectrum, given by \cite[Eq.~2.1.13]{ChewBook}  
\begin{align} \label{reflection_coeff} 
R(\kappa_x) &=  \frac{\kappa_{1z} - \kappa_{2z}}{\kappa_{1z} + \kappa_{2z}}
\end{align}
for dielectric materials;
this follows from \eqref{R_theta} after a change of variables to wavenumber coordinates while using \eqref{wavenumbers}.

Remarkably, a behavior of this sort characterizes a linear and space-invariant (LSI) system, which is fully described by its wavenumber response $R(\kappa_x)$ for any $\kappa_x$. 

\section{Plane Wave Spectral Representation}  \label{sec:Fourier_pw}

\begin{figure} [t!]
        \centering
	\includegraphics[width=.999\linewidth,tics=10]{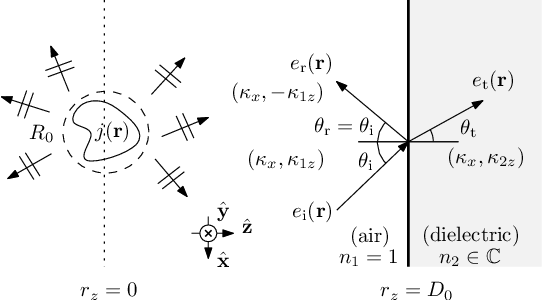} \vspace{-0.0cm}
                \caption{Scalar wave propagation in a 3D isotropic and inhomogeneous medium. View from the plane of incidence.} \vspace{0.0cm}
                \label{fig:Fig2} 
        \end{figure}

Consider now every possible vertical plane obtainable by rotating the $xz$-plane of incidence (i.e., $\phi_{\text i}=0$) about the $x$-axis by an angle ${\phi_{\text i} \in [0,2\pi)}$. This brings into play other variables in the spatial and wavenumber domains, which we embed into the vectors $\vect{r}$ with coordinates $(r_x,r_y)$ and $\vect{\kappa}$ with coordinates $(\kappa_x,\kappa_y)$.

The field $e_{\text i}(\vect{r},r_z)$ radiated by a source of electric current $j(\vect{r},r_z)$ is  described exactly by an integral superposition of complex harmonics of different amplitudes and spatial frequencies via the Fourier (plane wave) spectral representation \cite{ChewBook,PlaneWaveBook}.
Precisely, for a source enclosed within a sphere of radius ${0 < R_0 < D_0}$ (see Fig.~\ref{fig:Fig2}),
 \begin{align}   \label{incident_field}
e_{\text i}(\vect{r},r_z) &  =
\begin{cases} 
\begin{aligned} \displaystyle
& \iint_{-\infty}^{\infty} E_{\text i}^-(\vect{\kappa}) \, e^{\imagunit \vect{\kappa}^{\Ttran} \vect{r}}  \frac{d\vect{\kappa}}{(2\pi)^2}
& \; r_z < -R_0\\\displaystyle
& \iint_{-\infty}^{\infty}  E_{\text i}^+(\vect{\kappa}) \, e^{\imagunit \vect{\kappa}^{\Ttran} \vect{r}}  \frac{d\vect{\kappa}}{(2\pi)^2}
& r_z > R_0
\end{aligned}
\end{cases}
\end{align}
with complex-valued amplitudes
\begin{equation} \label{incident_spectrum}
E_{\text i}^\pm(\vect{\kappa}) =  \frac{\kappa_1 \eta_1}{2} \frac{J_\pm(\vect{\kappa})}{\kappa_{1z}} \, e^{\pm\imagunit  \kappa_{1z} r_z}
\end{equation}
specified by the source's spectrum $J_\pm(\vect{\kappa})$ obtained via a 3D Fourier transform of $j(\vect{r},r_z)$ evaluated at ${\kappa_z =  \pm \kappa_{1z}}$, $\kappa_{iz}$ being defined as
 \begin{equation} \label{kappaz_1} 
\kappa_{iz} =  \sqrt{\kappa_i^2 - \|\vect{\kappa}\|^2},
\end{equation}
for $i=1,2$. Thus,
\begin{equation} \label{source_spectrum}
J_\pm(\vect{\kappa}) \!=\! \iiint_{-\infty}^\infty  \!\! j(\vect{s},s_z) \, e^{-\imagunit \left(\vect{\kappa}^{\Ttran} \vect{s} \pm \kappa_{1z} s_z\right)} \, d\vect{s} ds_z
\end{equation}
given ${\eta_1 \approx 120 \pi}$ as the wave impedance of free-space.

The reflected field $e_{\text r}(\vect{r})$ follows from the linearity of the spatial filtering operation applied by the surface and the delay property of the Fourier transform, as the surface is placed at an arbitrary distance $D_0$ from the source, along the $z$-axis; see Fig.~\ref{fig:Fig2}.
The Fourier spectral representation of $e_{\text r}(\vect{r})$ is therefore
\begin{equation} \label{reflected_field_total}
e_{\text r}(\vect{r},r_z) = \!
 \iint_{-\infty}^\infty  E_{\text i}^+(\vect{\kappa})  R(\vect{\kappa}) \, e^{- \imagunit \kappa_{1z} (r_z-2D_0)}  e^{\imagunit \vect{\kappa}^{\! \Ttran} \! \vect{r}}  \frac{d\vect{\kappa}}{(2\pi)^2}
\end{equation}
with $R(\vect{\kappa})$ the Fresnel spectrum in \eqref{reflection_coeff} and $\kappa_{1z}$ as defined in \eqref{kappaz_1}.
Physically, the reflected field is created by superimposing the interactions with all possible incident contributions on the plane of incidence and for all possible vertical planes.
With respect to an incident plane wave, a reflected plane wave exhibits an extra phase shift that accounts for the round-trip delay accumulated by the incident wave during the travel 
to the surface and back, along the $z$-axis. This effect can be regarded as a \emph{migration} of the incident field and is directly connected to the image theorem, as discussed in Sec.~\ref{sec:image_th}.

\section{Image Theorem} \label{sec:image_th} 

 \begin{figure*}[th!]
\begin{equation} \tag{27} \label{received_field_total}
e(\vect{r},r_z) = 
\begin{cases}
\begin{aligned} \displaystyle 
& \iint_{-\infty}^\infty   \left(E_{\text i}^-(\vect{\kappa})  e^{-\imagunit  \kappa_{1z} r_z} + E_{\text i}^+(\vect{\kappa}) R(\vect{\kappa})  e^{- \imagunit \kappa_{1z} (r_z-2D_0)} \right)  e^{\imagunit \vect{\kappa}^{\Ttran} \vect{r}}  \frac{d\vect{\kappa}}{(2\pi)^2} & \quad r_z < -R_0 \\ \displaystyle 
& \iint_{-\infty}^\infty  E_{\text i}^+(\vect{\kappa}) \left( e^{\imagunit  \kappa_{1z} r_z} + R(\vect{\kappa}) e^{- \imagunit \kappa_{1z} (r_z-2D_0)} \right) e^{\imagunit \vect{\kappa}^{\Ttran} \vect{r}} \frac{d\vect{\kappa}}{(2\pi)^2} & \quad R_0 < r_z \le D_0 
\end{aligned}
\end{cases}
\end{equation}
\hrule
\vspace{.2cm}
 \begin{equation} \tag{30} \label{wavenumber_response}
H(\vect{\kappa}; r_z,s_z) = 
\begin{cases} 
\begin{aligned} \displaystyle 
& \frac{\kappa_1 \eta_1}{2}  \frac{\mathbbm{1}_{\mathcal{D}}(\vect{\kappa})}{\kappa_{1z}}  \left(e^{-\imagunit \kappa_{1z} (r_z-s_z)}+ R(\vect{\kappa}) e^{-\imagunit \kappa_{1z} (r_z+s_z - 2D_0)} \right) & \quad r_z < -R_0 \\ \displaystyle
& \frac{\kappa_1 \eta_1}{2}  \frac{\mathbbm{1}_{\mathcal{D}}(\vect{\kappa})}{\kappa_{1z}}   \left(e^{\imagunit \kappa_{1z} (r_z-s_z)}  +   R(\vect{\kappa})  e^{-\imagunit  \kappa_{1z} (r_z+s_z- 2D_0)} \right) & \quad R_0 < r_z \le D_0 
\end{aligned}
\end{cases}
\end{equation}
\hrule
\vspace{.2cm}
\begin{equation} \tag{34} \label{kernel_reflection}
\vect{H}(\vect{k},\vect{\kappa}) =  \frac{\kappa_1 \eta_1}{2} \delta(\vect{k}-\vect{\kappa}) 
 \frac{\mathbbm{1}_{\mathcal{D}}(\vect{\kappa})}{\kappa_{1z}}
\cdot
\begin{cases}
\begin{aligned}
&
\begin{pmatrix}
0& 0 \\
R(\vect{k}) e^{\imagunit  2 \kappa_{1z} D_0} &  1
\end{pmatrix} & r_z < -R_0 \\
&
\begin{pmatrix}
1& 0 \\
R(\vect{k}) e^{\imagunit  2 \kappa_{1z} D_0} & 0
\end{pmatrix}   & R_0 < r_z \le D_0
\end{aligned}
\end{cases}
\end{equation}
\hrule
\end{figure*}

Plugging \eqref{incident_spectrum} into \eqref{incident_field}, the incident field in $\{r_z >  R_0\}$ is
 \begin{align}   \label{incident_field_expanded}
e_{\text i}(\vect{r},r_z) \!=\!  \frac{\kappa_1 \eta_1}{2} \! \iint_{-\infty}^\infty  \!\! \frac{J_{+}(\vect{\kappa})}{\kappa_{1z}} 
\, e^{\imagunit \left( \vect{\kappa}^{\Ttran} \vect{r} + \kappa_{1z} r_z\right)} \frac{d\vect{\kappa}}{(2\pi)^2}
\end{align}
where $J_+(\vect{\kappa})$ is given by \eqref{source_spectrum}.
Similarly, the reflected field in \eqref{reflected_field_total} can be rewritten  as
 \begin{align}   \label{reflected_field_image}
e_{\text r}(\vect{r},r_z) =  \frac{\kappa_1 \eta_1}{2} \! \iint_{-\infty}^\infty  \! \frac{J_{\text r}(\vect{\kappa})}{\kappa_{1z}} 
\, e^{\imagunit \left( \vect{\kappa}^{\Ttran} \vect{r} - \kappa_{1z} r_z\right)} \frac{d\vect{\kappa}}{(2\pi)^2}
\end{align}
where 
\begin{align} \label{equivalent_source_reflection}
J_{\text r}(\vect{\kappa}) = J_+(\vect{\kappa}) \, e^{\imagunit \kappa_{1z} 2D_0} \, R(\vect{\kappa}) .
\end{align}
Notice that \eqref{reflected_field_image} and \eqref{incident_field_expanded} have the same form. 
Hence, $J_{\text r}(\vect{\kappa})$ may be regarded as the Fourier spectrum of a fictitious source $j_{\text r}(\vect{r},r_z)$.
For $R(\vect{\kappa}) = -1$, the reflected field in \eqref{reflected_field_image} may be reproduced by replicating the source at $r_z = 2D_0$, which accounts for the field migration to the surface and backward. This is the \emph{image theorem}, whereby the reflection elicited by a perfect conductor is equivalent to a mirror image of the source  \cite[Sec.~4.7.1]{BalanisBook}.
As an example, for a point source $j(\vect{r},r_z) = \delta(\vect{r}) \delta(r_z)$, i.e., for $J_+(\vect{\kappa})=1$, applying Weyl's identity \cite[Eq.~2.2.27]{ChewBook}
\begin{equation} \label{Weyl_identity}
\frac{e^{\imagunit \kappa_1 \|(\vect{r},|r_z|)\|}}{\|(\vect{r},|r_z|)\|}= \frac{\imagunit}{2\pi}  \iint_{-\infty}^{\infty} \frac{e^{\imagunit ( \vect{\kappa}^{\Ttran} \vect{r} + \kappa_{1z} |r_z|)}}{\kappa_{1z}}  \, d\vect{\kappa},
\end{equation}
from \eqref{reflected_field_image} we obtain
\begin{equation} \label{field_point_source}
 e_{\text r}(\vect{r},r_z) = \imagunit \frac{\kappa_1 \eta_1}{4\pi} G(\vect{r},r_z,\vect{0},2 D_0)
  \end{equation}
where 
\begin{equation} \label{Green}
G(\vect{r},r_z,\vect{r}^\prime,r_z^\prime) =  \frac{e^{\imagunit \kappa_1 \|(\vect{r}-\vect{r}^\prime,r_z-r_z^\prime)\|}}{4\pi \, \|(\vect{r}-\vect{r}^\prime,r_z-r_z^\prime)\|}
\end{equation}
is the Green's function
describing a spherical wave generated at $(\vect{r}^\prime,r_z^\prime)$ and measured at $(\vect{r},r_z)$. Hence, $j_{\text r}(\vect{r},r_z) = \delta(\vect{r}) \delta(r_z - 2 D_0)$.

For arbitrary materials, $j_{\text r}(\vect{r},r_z)$ is obtained from the spatial convolution 
\begin{equation} \label{equivalent_current_density}
j_{\text r}(\vect{r},r_z) = \iint_{-\infty}^\infty  j(\vect{u},r_z-2D_0) \, r(\vect{r}-\vect{u}) \, d\vect{u}
\end{equation}
of the image source and the impulse response of the surface, 
\begin{equation} \label{Fig3}
r(\vect{r}) = \iint_{-\infty}^{\infty} R(\vect{\kappa}) \, e^{\imagunit \vect{\kappa}^{\Ttran} \vect{r}} \frac{d\vect{\kappa}}{(2\pi)^2},
\end{equation}
which is defined as the 2D inverse Fourier transform of $R(\vect{\kappa})$ in \eqref{reflection_coeff}. The azimuthal dependance of $r(\vect{r})$ can  be eliminated by evaluating \eqref{Fig3} at $(\|\vect{r}\|,0)$, which is possible due to the circular symmetry of $R(\vect{\kappa})$.

From~\eqref{equivalent_current_density}, we infer that the spatial filtering applied by the surface creates a \emph{blurred image} of the source. This effect vanishes in perfect conductors, recreating a perfect image.
For a point source, $j_{\text r}(\vect{r},r_z) = r(\vect{r}) \delta(r_z - 2D_0)$, $\vect{r}\in \Real^2$, modeling the impressed currents induced by the source on the entire surface.

The spatial filtering simplifies when the surface is far enough from the source that the reflected propagation occurs in the \emph{paraxial regime}. Then, $R(\vect{\kappa})$ is roughly constant for all possible incident angles and given by the complex material reflectivity \cite[Sec.~1.5.3]{OpticsBook}.
Due to the impulsiveness of the reflection mechanism under the paraxial assumption, the image source becomes a weakened (and phase-shifted) version of the original one, which is the premise of ray-tracing algorithms.
However, this need not be the case in wide-aperture MIMO, which rests on the range being short; this aspect is further expounded in Sec.~\ref{sec:ray_tracing}. 
The implications 
on the optimization of antenna spacings in MIMO communication are discussed in Sec.~\ref{sec:numerical_results}.

\section{Channel Impulse Response} \label{sec:channel_response}


A complete description of what unfolds in region~1 is obtained by combining all contributions into
\begin{equation} \label{total_field}
e(\vect{r},r_z) = e_{\text i}(\vect{r},r_z) + e_{\text r}(\vect{r},r_z)
\end{equation}
whose expression is given by \eqref{received_field_total} after substituting \eqref{incident_field} and \eqref{reflected_field_total}.
The input-output relationship between $j(\vect{s},s_z)$ and $e(\vect{r},r_z)$ is the spatial convolution \cite{PizzoIT21}
\setcounter{equation}{27}
\begin{equation}
e(\vect{r},r_z) = \iiint_{-\infty}^\infty j(\vect{s},s_z) \, h(\vect{r},r_z,\vect{s},s_z) \, d\vect{s} ds_z
\end{equation}
where $h(\vect{r},r_z,\vect{s},s_z)$ is the channel impulse response.
Combining \eqref{received_field_total}, \eqref{incident_spectrum}, and \eqref{source_spectrum}, the channel response can be written as the 2D inverse Fourier transform 
\begin{equation} \label{impulse_response_Fourier}
h(\vect{r}-\vect{s};r_z,s_z) = \iint_{-\infty}^{\infty}   H(\vect{\kappa};r_z,s_z) \, e^{\imagunit \vect{\kappa}^{\Ttran} (\vect{r}-\vect{s})} \, \frac{d\vect{\kappa}}{(2\pi)^2}
\end{equation}
of $H(\vect{\kappa};r_z,s_z)$ in \eqref{wavenumber_response}.
Here, the integration domain is practically limited to a disk $\mathcal{D}$ of radius $\kappa_1 =2\pi/\lambda$, correctly showing the low-pass-filtering behavior of the wireless propagation \cite{PizzoTSP21,PizzoIT21}, which is then converted into a functional dependence through an indicator function.
The reflected channel is space invariant over any pair of parallel $z$-planes. This extends to any  pair of parallel planes, not necessarily $z$, for an LOS channel. 

The space invariance is a direct consequence of the unboundedness and smoothness of the reflecting surface and enables a linear-system-theoretic interpretation of the reflection and transmission phenomena. Precisely, communications between any two different $z$-planes cutting source and receiver can be regarded as an LSI system with the wavenumber response in  \eqref{wavenumber_response}. There are three main terms in \eqref{wavenumber_response}, plus a phase shift due to migration, that may be interpreted as the cascade of:
\begin{itemize}
\item First, $\mathbbm{1}_{\mathcal{D}}(\vect{\kappa})$, a low-pass filter introduced by the migration operation \cite{PizzoIT21,PizzoTSP21}. 
\item Then, $1/\kappa_{1z}$, which confers a reverse-bowl behavior to $H(\vect{\kappa};r_z,s_z)$ and is directly attributable to the wave equation \cite{PizzoJSAC20,PizzoIT21}.
\item Finally, $R(\vect{\kappa})$ models the reflection. This depends on $\vect{\kappa}$ via $\kappa_{iz}$ in \eqref{kappaz_1}, hence it is circularly symmetric in the wavenumber domain, which is instrumental to devise an efficient numerical procedure to generate channel samples (see Appendix).
\end{itemize}


The space-invariant channel in \eqref{impulse_response_Fourier} generated by a specular reflection is obtainable as a particular instance of the double 2D Fourier transform \cite[Sec.~III]{PizzoIT21} 
\setcounter{equation}{30}
\begin{align} \notag
& h(\vect{r},r_z,\vect{s},s_z) = 
 \iiiint_{-\infty}^\infty  H(\vect{k},\vect{\kappa};r_z,s_z)  \\&\hspace{3cm}   \label{channel_response_complete}
 \cdot e^{\imagunit \vect{k}^{\Ttran}\vect{r}} e^{-\imagunit \vect{\kappa}^{\Ttran}\vect{s}}  \, \frac{d\vect{k}}{(2\pi)^2} \frac{d\vect{\kappa}}{(2\pi)^2}  
\end{align}
of the wavenumber response
\begin{equation} \label{angular_response_complete}
H(\vect{k},\vect{\kappa};r_z,s_z) = 
\vect{\phi}^{\Htran}(\vect{k},r_z)  
\vect{H}(\vect{k},\vect{\kappa})
\vect{\phi}(\vect{\kappa},s_z),
\end{equation}
given $\vect{\phi}(\vect{\kappa},s_z) = \left(e^{-\imagunit \kappa_{1z} s_z},e^{\imagunit \kappa_{1z} s_z}\right)^{\Ttran}$.
The above is parametrized by the wavenumber matrix 
\begin{align} \label{kernel_matrix}
\vect{H}(\vect{k},\vect{\kappa}) = 
\begin{pmatrix}
H_{++}(\vect{k},\vect{\kappa})& H_{+-}(\vect{k},\vect{\kappa}) \\
H_{-+}(\vect{k},\vect{\kappa}) & H_{--}(\vect{k},\vect{\kappa})
\end{pmatrix} 
\end{align}
that models the coupling between every input spatial frequency $\vect{\kappa}$ and every other output spatial frequency $\vect{k}$.
It can also be regarded as an angular response 
mapping every incident plane wave traveling along $(\vect{\kappa},\pm \kappa_{1z})$ into every other receive plane wave from $(\vect{k},\pm \kappa_{1z})$. The convention adopted for the entries of \eqref{kernel_matrix} is that the first and second subscripts refer, respectively, to received and incident plane waves (each one being associated with upgoing or downgoing waves).

We next find the parameterization of $\vect{H}(\vect{k},\vect{\kappa})$ that models the scenario in Sec.~\ref{sec:Fourier_pw}. By inspection, comparing \eqref{impulse_response_Fourier}--\eqref{wavenumber_response} against \eqref{channel_response_complete}, yields \eqref{kernel_reflection}. The entries of the angular matrix are impulsive because incident and received plane waves are in one-to-one correspondence: each incident wave turns into a received wave with specular direction, as specified by Snell's law.

Generally, the surface of a material object may appear as either smooth or rough depending on the frequency. A rough surface at microscopic level reflects every impinging plane wave off multiple directions creating a diffuse reflection spectrum, typically centered around the specular direction. 
These surface irregularities are accounted by a non-impulsive $\vect{H}(\vect{k},\vect{\kappa})$ in \eqref{angular_response_complete}, whose computation is left for future work.

\section{Application to MIMO Communication} \label{sec:numerical_results} 

  \begin{figure}
\centering\vspace{-0.0cm}
\includegraphics[width=.999\linewidth]{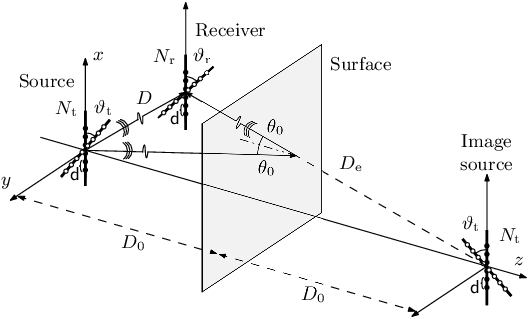} 
\caption{ULAs separated by $D$ and equipped with $N_{\text t} = N_{\text r}=8$ antennas with spacing ${\sf d}$. Arrays have arbitrary orientations $\vartheta_{\text t}$ and $\vartheta_{\text r}$ with respect to the $x$-axis.
The clear and solid circles at source and receiver indicate the antennas and their projections onto the $x$-axis, respectively.
Antennas are connected either via a LOS or a reflected channel off a $z$-oriented surface. We denote by $\theta_0$ the angle formed by the surface normal and the geometrical path connecting the centroids of the image source and receiver.}
\label{fig:Fig3}
\end{figure}

Let us now apply the developed model to evaluate the channel eigenvalues, DOF, and spectral efficiency. 

With $N_{\text t}$ transmit and $N_{\text r}$ receive antennas, the channel matrix $\vect{H} \in\Complex^{N_{\text r} \times N_{\text t}}$ is obtained by sampling the impulse response at the antenna locations, ${[\vect{H}]_{m,n} = h(\vect{r}_m,\vect{s}_n)}$ for $m=0, \dots, N_{\text r}-1$ and $n=0, \dots, N_{\text t}-1$. The transmit array is centered at the origin whereas the centroid of the receive array is at $\vect{r}_0 = (r_{0x},r_{0y},r_{0z})$.
An efficient numerical generation procedure for $\vect{H}$ is provided in the Appendix.

Let $N_{\min} = \min(N_{\text r}, N_{\text t})$ and $N_{\max} = \max(N_{\text r}, N_{\text t})$.
We consider uniform linear arrays (ULAs) at $57.5$~GHz (see Fig.~\ref{fig:Fig3}) under the proviso that those ULAs are substantially shorter than their separation range, the so-called \emph{paraxial approximation}, so we can leverage results available for LOS channels \cite{HeedongISIT,Heedong2021}.
The transmitting and receiving ULAs have arbitrary orientations $\vartheta_{\text t}$ and $\vartheta_{\text r}$ with respect to the $x$-axis. 

We hasten to emphasize that the reliance on the paraxial approximation is confined to the production of benchmark results for LOS MIMO, with our channel model being valid regardless. 
The frequency, in turn, is motivated by mmWave applications \cite{Sundeep2014} and by availability of refractive indices for most common materials \cite{Sato}.

\subsection{Parallel Arrays Optimized for LOS Transmission}

Consider parallel ULAs aligned with the $x$-axis, with ${N_{\text t} = N_{\text r}=8}$ and antenna spacing ${\sf d}$.
The range is ${D = 10}$~m whereas the surface is at ${D_0= 15}$~m.
First, we validate the model in LOS, for which the closed-form 
solution 
 in \eqref{Green} is available.
The channel matrix obtained by sampling \eqref{Green} is compared to the LOS component in our model, derivable after an inverse Fourier transform of the first term in \eqref{wavenumber_response}, the LOS term, according to \eqref{impulse_response_Fourier}, followed by spatial sampling.
Setting
\setcounter{equation}{34}
\begin{equation} \label{Rayleigh}
{\sf d}(D)=\sqrt{\lambda D/N_{\max}},
\end{equation}
renders $\vect{H}$ a Fourier matrix and is optimum at high SNR \cite{Rodwell2011,HeedongISIT}. 
The normalized eigenvalues of $\vect{H} \vect{H}^{\Htran}$, $\lambda_n(\vect{H})$, are plotted in Fig.~\ref{fig:Fig4}. 
The perfect match validates the numerical procedure in the Appendix for this LOS setting. 

Then, we validate the model under perfect reflection. To this end, the channel obtained by imaging the source is compared against the one associated with the perfect reflection in our model; the latter is obtained by plugging the second term in \eqref{wavenumber_response} with $R(\vect{\kappa}) = -1$ into \eqref{impulse_response_Fourier} and sampling.

The eigenvalues of the reflected channel matrix are further shown in Fig.~\ref{fig:Fig4} for different materials.
These undergo two effects relative to their LOS brethren:
\begin{itemize}
 \item \emph{Power loss} caused by the longer range and by the reflection of only a share of the incident power, with dense materials and shallow angles reflecting better. 
 \item \emph{Spatial selectivity} due the antenna spacing in \eqref{Rayleigh} being suboptimally small  
 for the longer range of the reflected channel.
 \end{itemize}


\begin{figure}
        \centering
	\includegraphics[width=.999\columnwidth,tics=10]{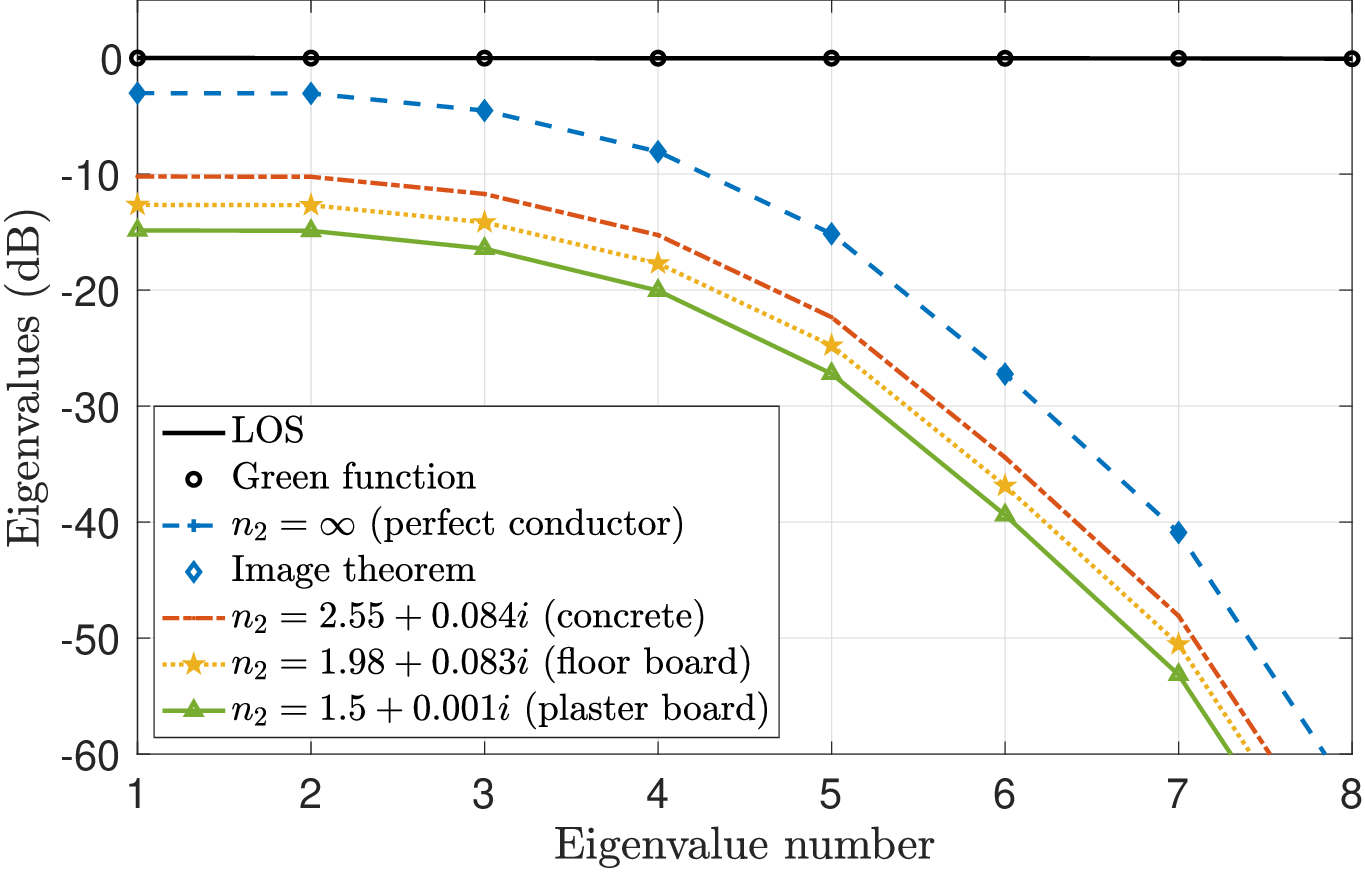} \vspace{-0.0cm}
                \caption{Normalized channel eigenvalues for various materials. Parallel ULAs 
              separated by $D=10$~m with spacing ${\sf d}(D)$ in \eqref{Rayleigh}.}
                \vspace{0.0cm}
                \label{fig:Fig4} 
        \end{figure}
        
\begin{figure}
\centering\vspace{-0.0cm}
\includegraphics[width=.999\linewidth]{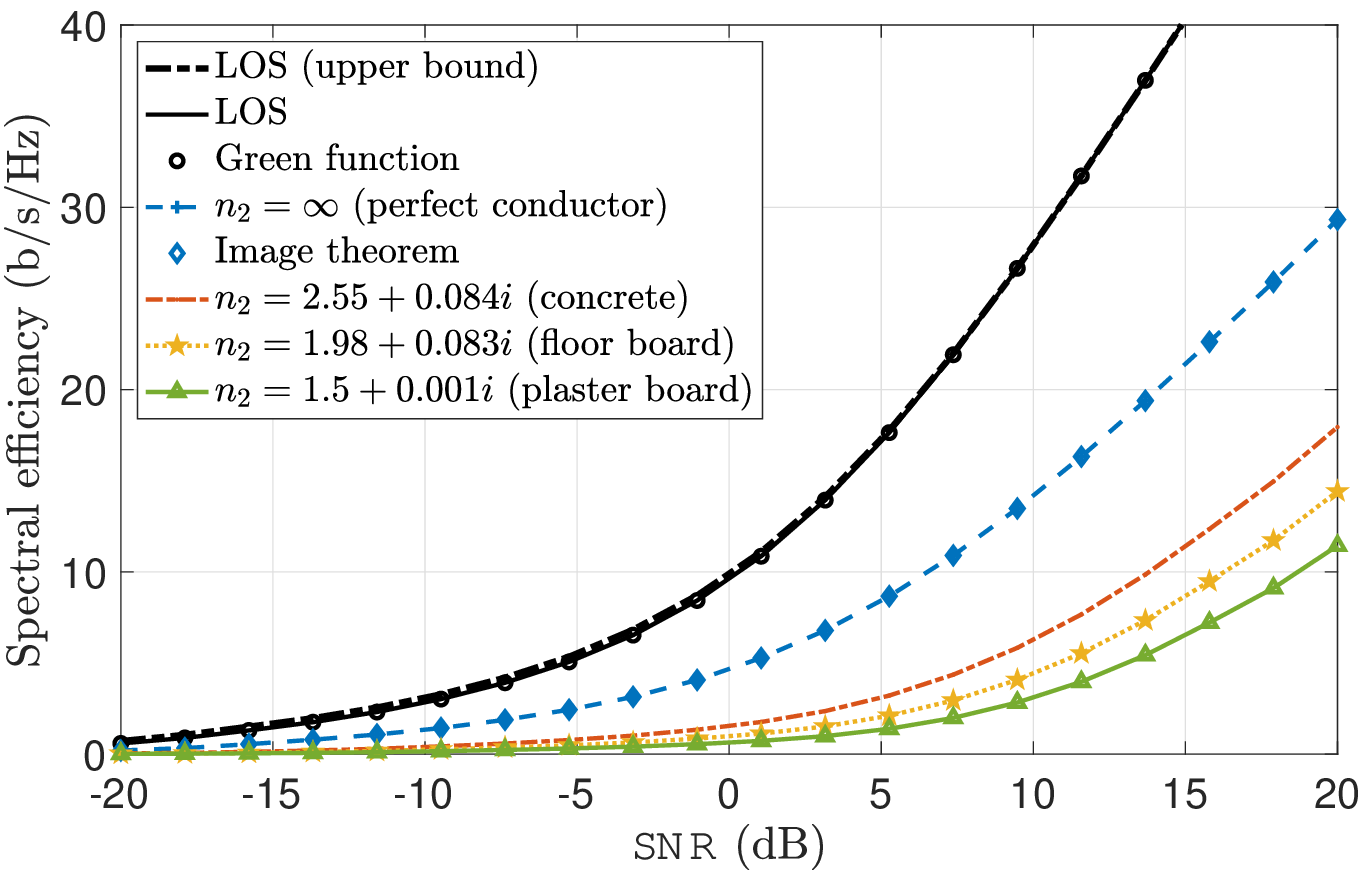} \vspace{-0.0cm}
\caption{Spectral efficiency as a function of SNR for various materials. Parallel ULAs 
separated by $D=10$~m with spacing ${\sf d}(D,\mathsf{SNR})$ in \eqref{spacing}.
}\vspace{-0.0cm}
\label{fig:Fig5}
\end{figure}

We now gauge the capacity 
with channel-state information at the transmitter, which
equals \cite{tulino2004mimo,LozanoBook}
\begin{align} \label{capacity}
C(\vect{H},{\sf SNR}) =\sum_{n=1}^{N_{\min}} \log_2 \! \left(1 +  \left(\nu - \frac{1}{\lambda_n(\vect{H})} \right)^{\!\!+} \lambda_n(\vect{H}) \! \right)
\end{align}
where $\nu$ is such that ${\sum_{n=1}^{N_{\min}} \left(\nu - {1}/{\lambda_n(\vect{H})} \right)^+ = {\sf SNR}}$ while ${\sum_{n=1}^{N_{\min}} \lambda_n(\vect{H}) = N_{\text r} N_{\text t}}$.
At any given SNR, $C({\sf SNR}) = \max_{\vect{H}} C(\vect{H},{\sf SNR})$ satisfies \cite{HeedongISIT,Heedong2021}
\begin{equation} \label{upper_bound}
C({\sf SNR}) \le \max_{\rho \in\{1,2, \ldots, N_{\min}\}} \rho\log_2 \! \left(1 + \frac{{\sf SNR}}{\rho}  \frac{N_{\text r} N_{\text t}}{\rho} \right) ,
\end{equation}
with the upper bound corresponding to $\rho$ nonzero eigenvalues equal to $N_{\text r} N_{\text t}/\rho$ and to the SNR-dependent antenna spacing
\begin{equation} \label{spacing}
{\sf d}(D,{\sf SNR}) = \sqrt{\eta({\sf SNR}) \lambda D/N_{\max}} ,
\end{equation} 
for a fraction $\eta({\sf SNR}) = \rho({\sf SNR})/N_{\min}$ of the $N_{\min}$ potential  DOF. 
Thus, $\eta\in[0,1]$ with $\eta=1$ at high enough SNR.
The capacity $C(\vect{H},{\sf SNR})$  is reported in Fig.~\ref{fig:Fig5} for the antenna spacing, ${\sf d}(D,{\sf SNR})$, that is optimum for the LOS channel at every SNR.
With respect to the LOS case, the capacity of the reflected channel experiences an offset (power loss, due to the longer range) and a reduced slope (DOF loss, due to the spatial selectivity).

\subsection{Parallel Arrays Optimized for the Reflected Transmission} \label{sec:MIMO_reflected}

 \begin{figure} [t!]
        \centering
	\includegraphics[width=.999\columnwidth,tics=10]{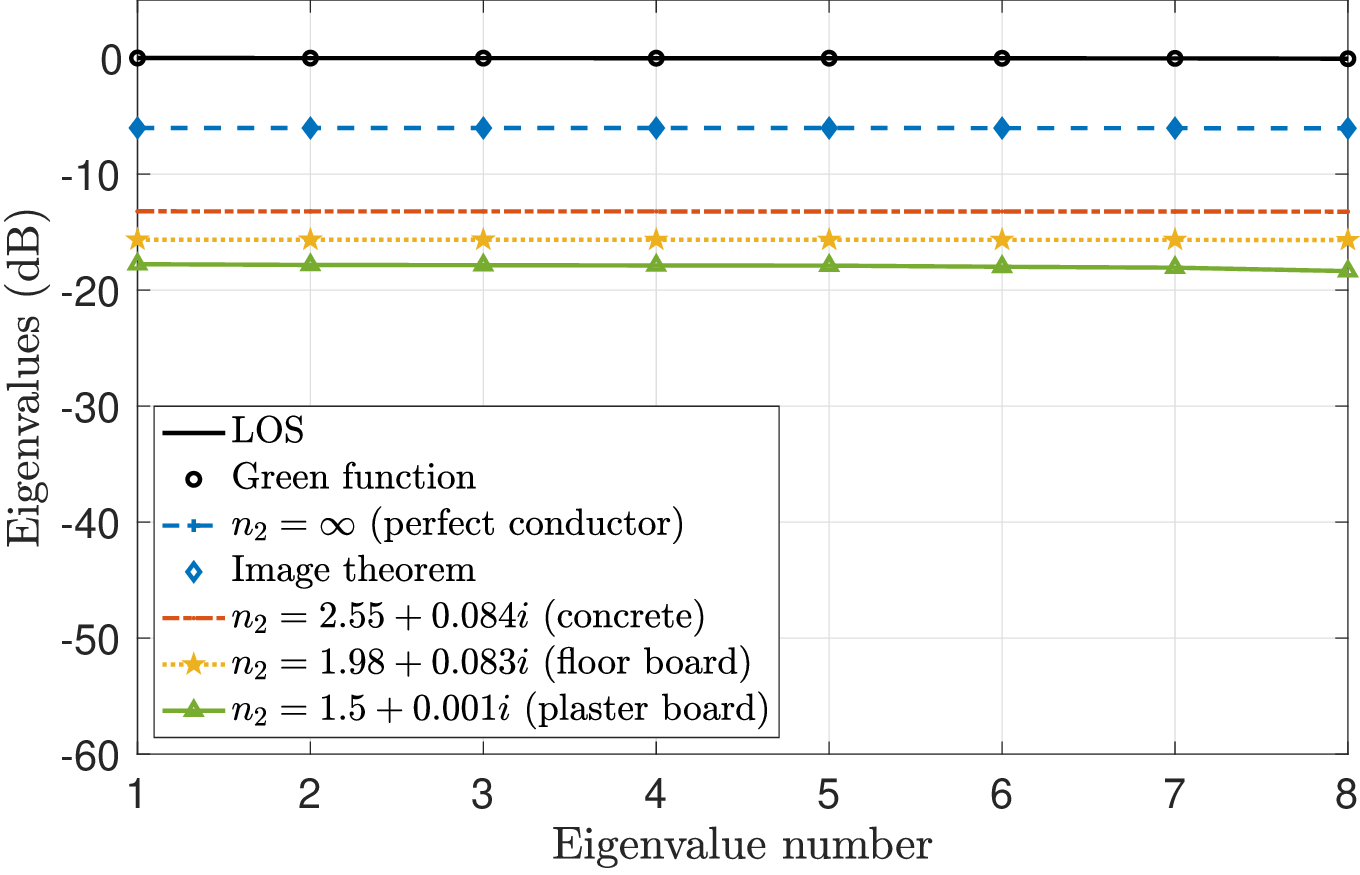} \vspace{-0.0cm}
                \caption{Normalized channel eigenvalues for various materials. Parallel ULAs 
                separated by $D=10$~m with spacing ${\sf d}(D)$ for the LOS channel and ${\sf d}(D_{\text e})$ for the reflected channel.}
                \vspace{0.0cm}
                \label{fig:Fig6} 
        \end{figure}     

\begin{figure}
\centering\vspace{-0.0cm}
\includegraphics[width=.999\linewidth]{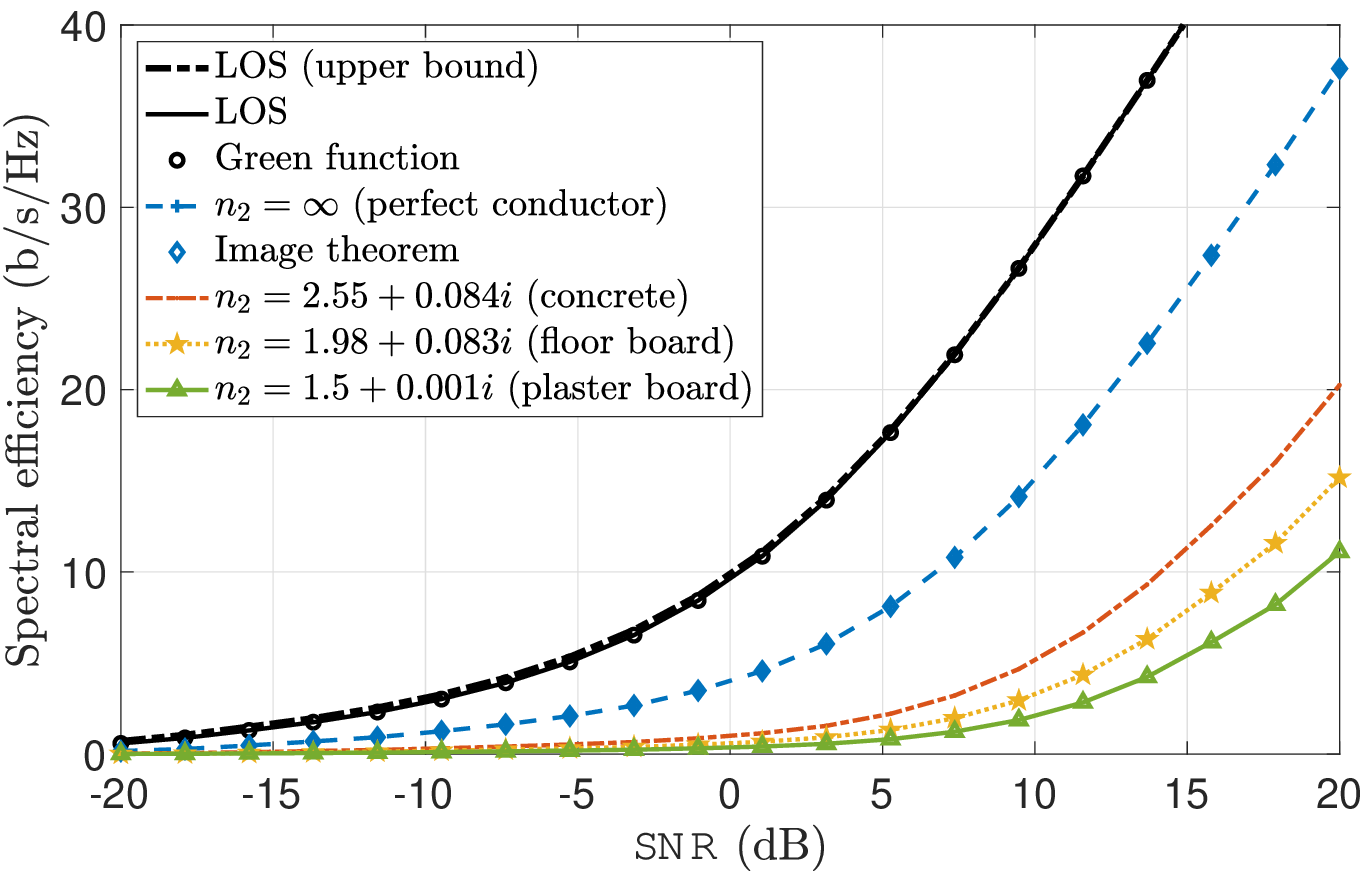} \vspace{-0.0cm}
\caption{Spectral efficiency as a function of SNR for various materials. Parallel ULAs 
separated by $D=10$~m with spacing ${\sf d}(D,{\sf SNR})$ for the LOS channel and ${\sf d}(D_{\text e},\mathsf{SNR})$ for the reflected channel.
}
\vspace{-0.0cm}
\label{fig:Fig7}
\end{figure}

 While the power loss is inevitable, because of the longer range, the spatial selectivity can be corrected by tailoring the antenna spacing to the equivalent LOS transmission from the image source. To this end, recall from the image theorem that the reflected channel can be regarded as an LOS channel with augmented distance $D_{\text e} > D$;
  in the setting of Figs.~\ref{fig:Fig4} and~\ref{fig:Fig5}, ${D_{\text e}= 2 D_0 - D}$. For a perfect conductor, this alone justifies the choice of an antenna spacing equal to ${\sf d}(D_{\text e})$. 
The argument is somewhat more involved for arbitrary materials, due to the distortion introduced by reflection, but it ultimately leads to the same observation as illustrated in Fig.~\ref{fig:Fig6}. 
Numerically, this is supported by the invariance of the curves for the materials in Fig.~\ref{fig:Fig4}. Physically, it is explained by the paraxial approximation, whereby the field has an approximately constant wavenumber response in magnitude. Hence, the reflection has an approximately multiplicative effect on the channel impulse response in~\eqref{wavenumber_response} and the whole interaction phenomenon with the surface is described by the reflectivity coefficient, $R(\theta_0)$, which is derivable from \eqref{R_theta} after setting ${\theta_{\text i}=\theta_0}$ with $\theta_{0}$ as per Fig.~\ref{fig:Fig3}.

Similarly, the eigenvalues of the reflected MIMO channel $\vect{H} \vect{H}^{\Htran}$ are obtained by scaling the LOS eigenvalues uniformly by $|R(\theta_0)|^2$.
From~\eqref{R_theta}, for the chosen materials, setting ${\theta_{\text i}=\theta_0}$ yields a scaling of $7.19$~dB (concrete), $9.63$~dB (floorboard), and $13.98$~dB (plaster board). These values describe the gap in Fig.~\ref{fig:Fig6} between the eigenvalues of the reflected channel for various materials and those of a perfect conductor. 
The additional gap to the LOS channel is due to the enhanced range, 
  a loss of $6.02$~dB in our setting.

For completeness, Fig.~\ref{fig:Fig7} shows the spectral efficiency corresponding to the eigenvalue distributions in Fig.~\ref{fig:Fig6}. With respect to Fig.~\ref{fig:Fig5}, the antenna spacing is 
${\sf d}(D,{\sf SNR})$ for the LOS channel and ${\sf d}(D_{\text e},{\sf SNR})$ for the reflected channel, which lead to the same DOF.

\subsection{Power Loss and Spatial Selectivity for Parallel Arrays}

\begin{figure} [t!]
        \centering
	\includegraphics[width=.999\columnwidth,tics=10]{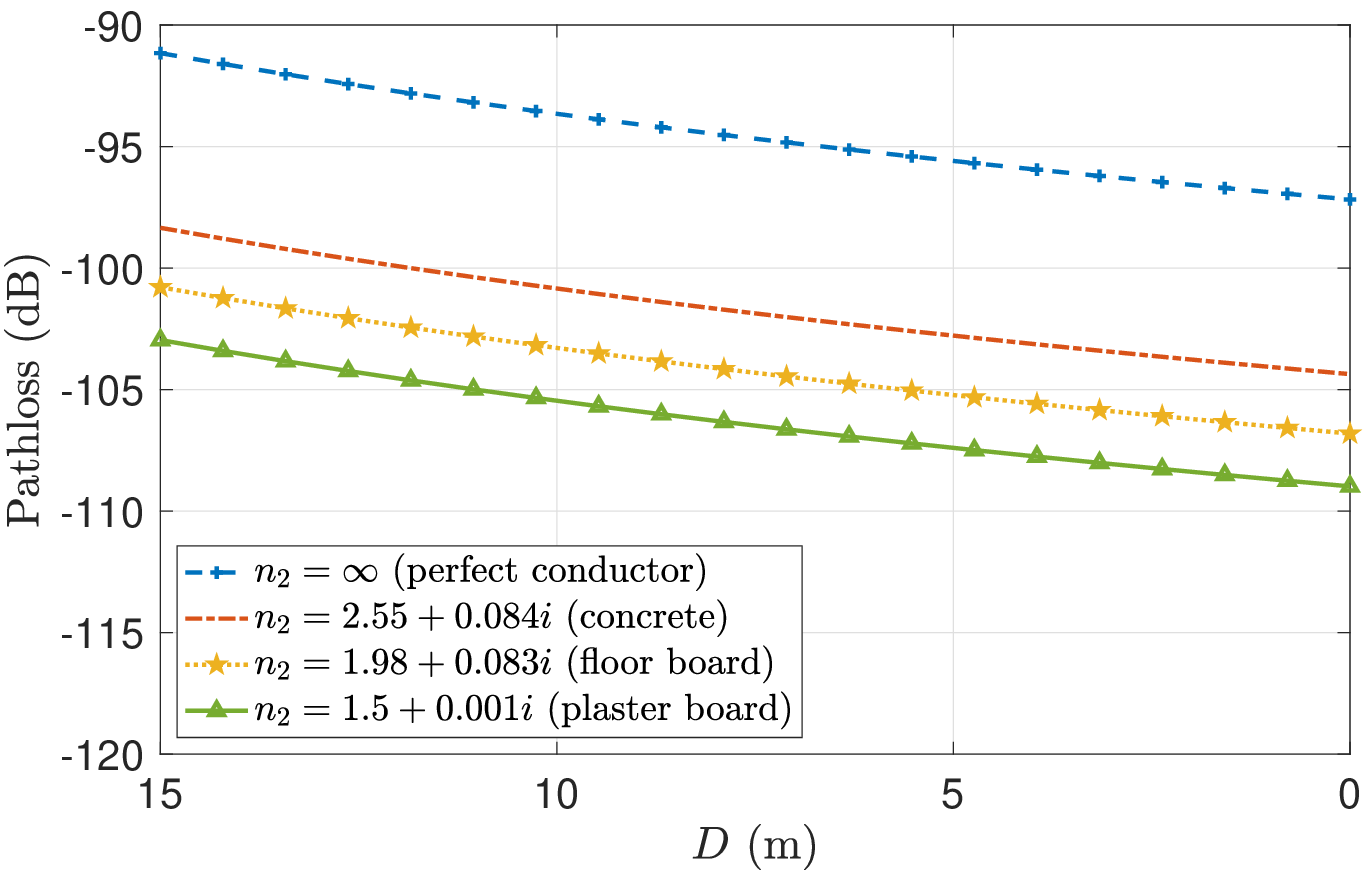} \vspace{-0.0cm}
                \caption{Pathloss as a function of $D$ for different materials at normal incidence.}
                \vspace{0.0cm}
                \label{fig:Fig8} 
        \end{figure}     
        

We have seen that the power loss is determined by the additional range and by the share of incident power not reflected by the surface. 
This is constant over the arrays themselves as amplitude variations thereon are negligible with the proviso that propagation occurs in the paraxial regime.
From the image theorem,
\begin{equation} \label{pathloss_0}
\beta = |R(0)|^2 \left(\frac{\lambda}{4 \pi D_{\text e}}\right)^2
\end{equation}
where $D_{\text e} = 2D_0 - D$ and $R(0) = (1 -  n_2)/(1 + n_2)$.
In Fig.~\ref{fig:Fig8}, $\beta$ is plotted as a function of $(D_0-D)$ for different materials. The interface is at $D_0 = 15$~m from the source, while the range between receiver and surface varies accordingly to $(D_0-D)$.

\begin{figure}
\centering\vspace{-0.0cm}
\includegraphics[width=.999\linewidth]{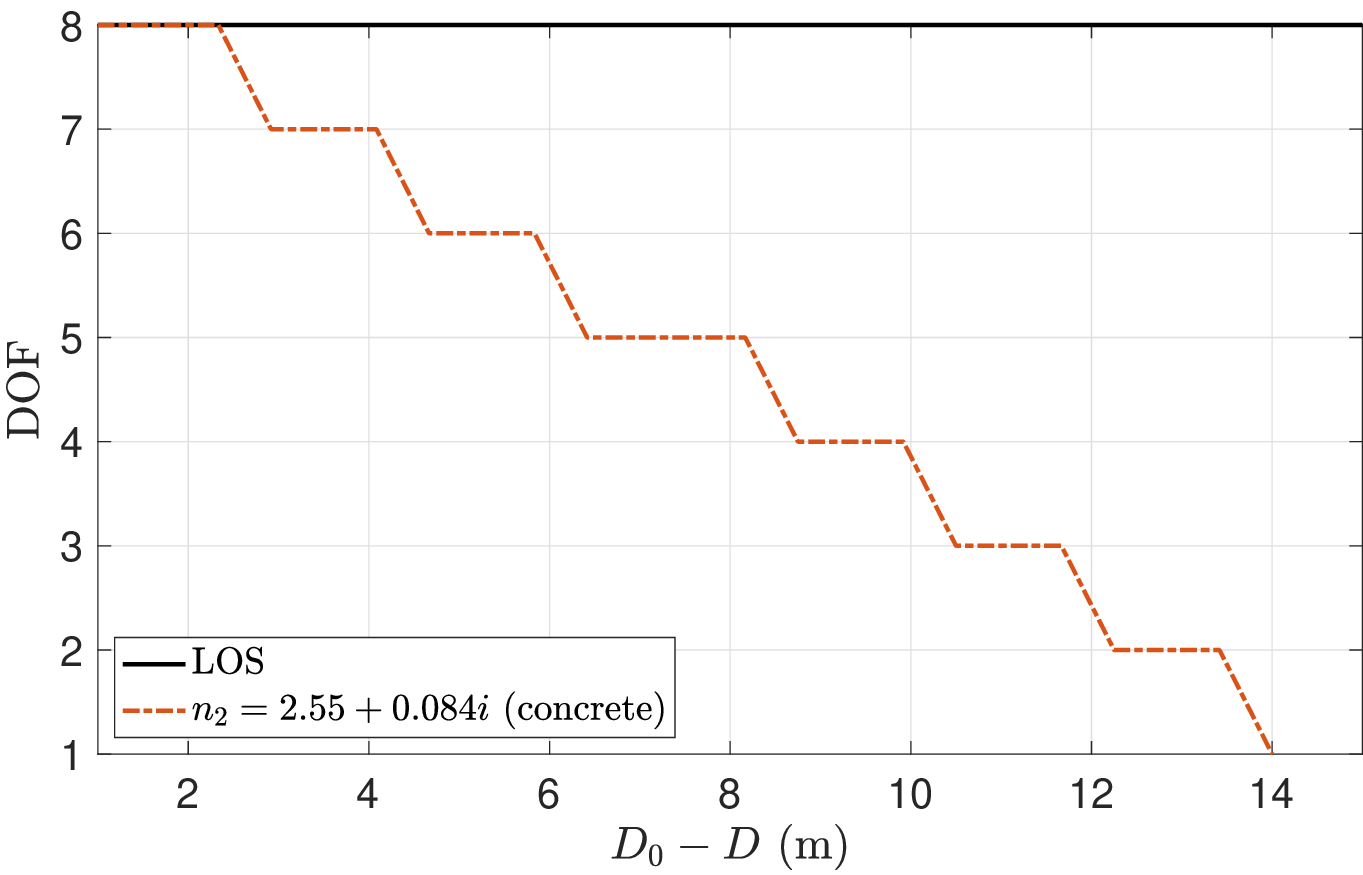} \vspace{-0.0cm}
\caption{Number of DOF as a function of $(D_0 - D)$ when the material is concrete. Parallel ULAs.}
\label{fig:Fig9}
\end{figure}

Receiver motion away from the surface, if unaccounted for, leads to a decreasing stepwise function of $D-D_0\in[0,D_0]$; this is shown in Fig.~\ref{fig:Fig9}, where the DOF equal the number of eigenvalues that are at most $40$~dB below the maximum.
Correcting the antenna spacing as a function of $D$ prevents this decrease. 

\subsection{Non-Parallel Arrays}

 \begin{figure} [t!]
        \centering
	\includegraphics[width=.999\columnwidth,tics=10]{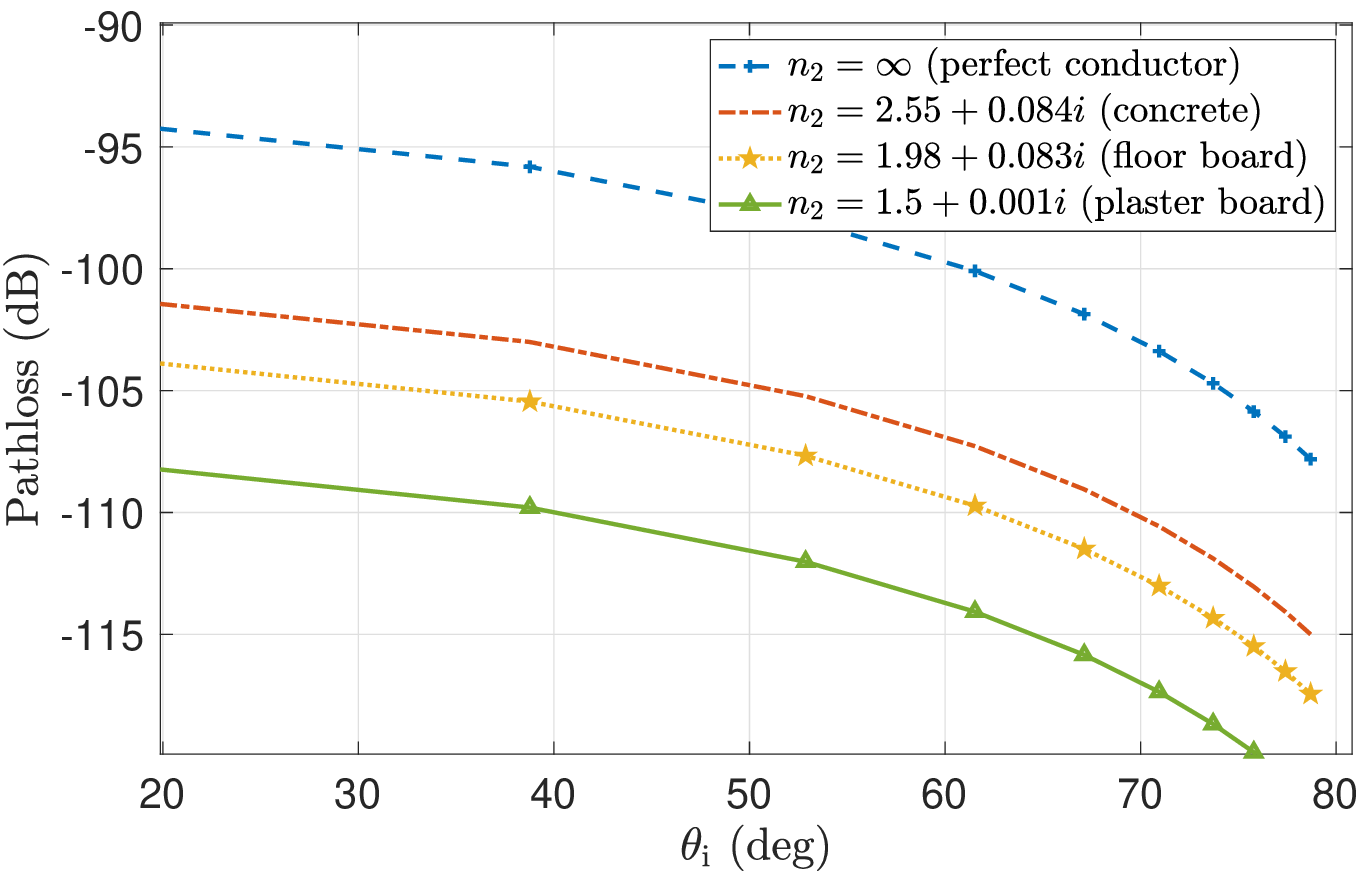}\vspace{-0.0cm}
                \caption{Pathloss as a function of $\theta_0$ for various materials. Oblique incidence with the receiver at $r_{0x}\in[0,100]$~m and $r_{0z}=10$~m.}
                \vspace{0.0cm}
                \label{fig:Fig10} 
        \end{figure}    

Non-parallel ULA configurations arise either when the receiver is shifted along the $x$-axis, creating an oblique incidence ($\theta_0 > 0$), or when arrays are oriented differently in elevation ($\vartheta_{\text t} \ne \vartheta_{\text r}$); see Fig.~\ref{fig:Fig3}. 
The relative azimuth angle is set to zero, as it is immaterial to ULAs \cite{Heedong2021}.
With the focus on oblique incidence and its impact on power loss and spatial selectivity, the ULAs are aligned with the $x$-axis ($\vartheta_{\text t} = \vartheta_{\text r} = 0$). 

First, let us consider the power loss.
Due to rotational symmetry about the $x$-axis, the $xz$-plane can be selected without loss of generality.
The pathloss in \eqref{pathloss_0} generalizes to arbitrary receive positions when using
\begin{equation} \label{equivalent_dist_theta}
D_{\text e}(\theta_0) = \frac{2 D_0-r_{0z}}{\cos(\theta_0)},
\end{equation}
and $R(\theta_0)$, which 
are parametrized by the incident angle
\begin{equation} \label{theta_incident}
\theta_0 = \arccos \! \left(\frac{2 D_0 - r_{0z}}{\sqrt{D^2 + 4 D_0 (D_0-r_{0z})}}\right).
\end{equation}
Fig.~\ref{fig:Fig10} depicts $\beta$ for various materials. 
The receiver is shifted along the $x$-axis on the interval $r_{0x}\in[0,100]$~m with $r_{0z}=10$~m such that $D=(r_{0x}^2+r_{0z}^2)^{1/2}$. 

 \begin{figure}[t!]
\begin{subfigure}{\columnwidth}
\includegraphics[width=.999\linewidth]{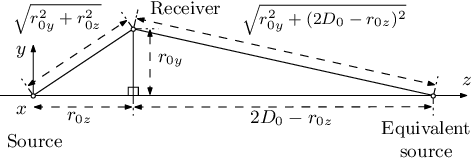}
  \caption{Projected view on the $yz$-plane (top view of Fig.~\ref{fig:Fig3}).}\label{fig:Fig11_top}
\end{subfigure}
\begin{subfigure}{\columnwidth}
\vspace{.3cm}
  \includegraphics[width=.999\linewidth]{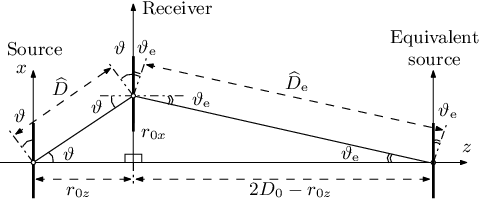}
  \caption{Projected view on the $xz$-plane (side view of Fig.~\ref{fig:Fig3}).}\label{fig:Fig11_side}
\end{subfigure}
\begin{subfigure}{\columnwidth}
\vspace{.3cm}
  \includegraphics[width=.999\linewidth]{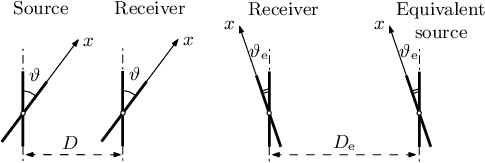}
  \caption{Modeling of oblique incidence as a relative orientation.}\label{fig:Fig11_model}
\end{subfigure}
\caption{Non-parallel ULA configuration arising from an oblique incidence with ULAs oriented as the $x$-axis.}
\label{fig:Fig11}
\end{figure}

    \begin{figure} [t!]
        \centering
	\includegraphics[width=.999\columnwidth,tics=10]{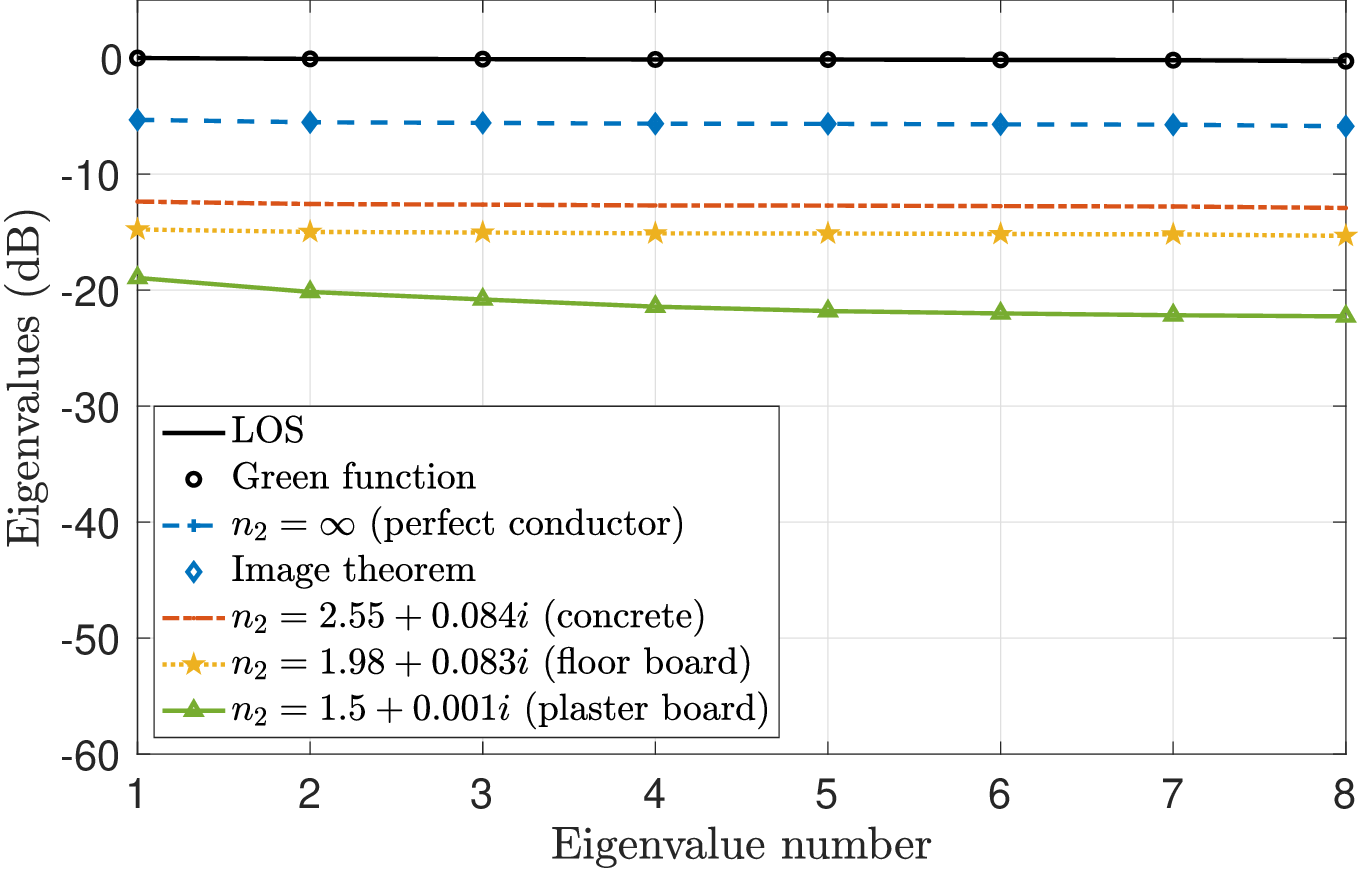} \vspace{-0.0cm}
                \caption{Normalized channel eigenvalues for various materials. Oblique incidence with the receiving ULA at  $\vect{r}_0 = (1,4,10)$~m (hence, $\vartheta=5.3^\circ$ and $\vartheta_{\text e}=2.8^\circ$).
                The antenna spacings are ${\sf d}(D,{\sf SNR},\vartheta)$ for the LOS channel and ${\sf d}(D_{\text e},{\sf SNR},\vartheta_{\text e})$ for the reflected channel.}
                \vspace{0.0cm}
                \label{fig:Fig12} 
        \end{figure}  

\begin{figure}
\centering\vspace{-0.0cm}
\includegraphics[width=.999\linewidth]{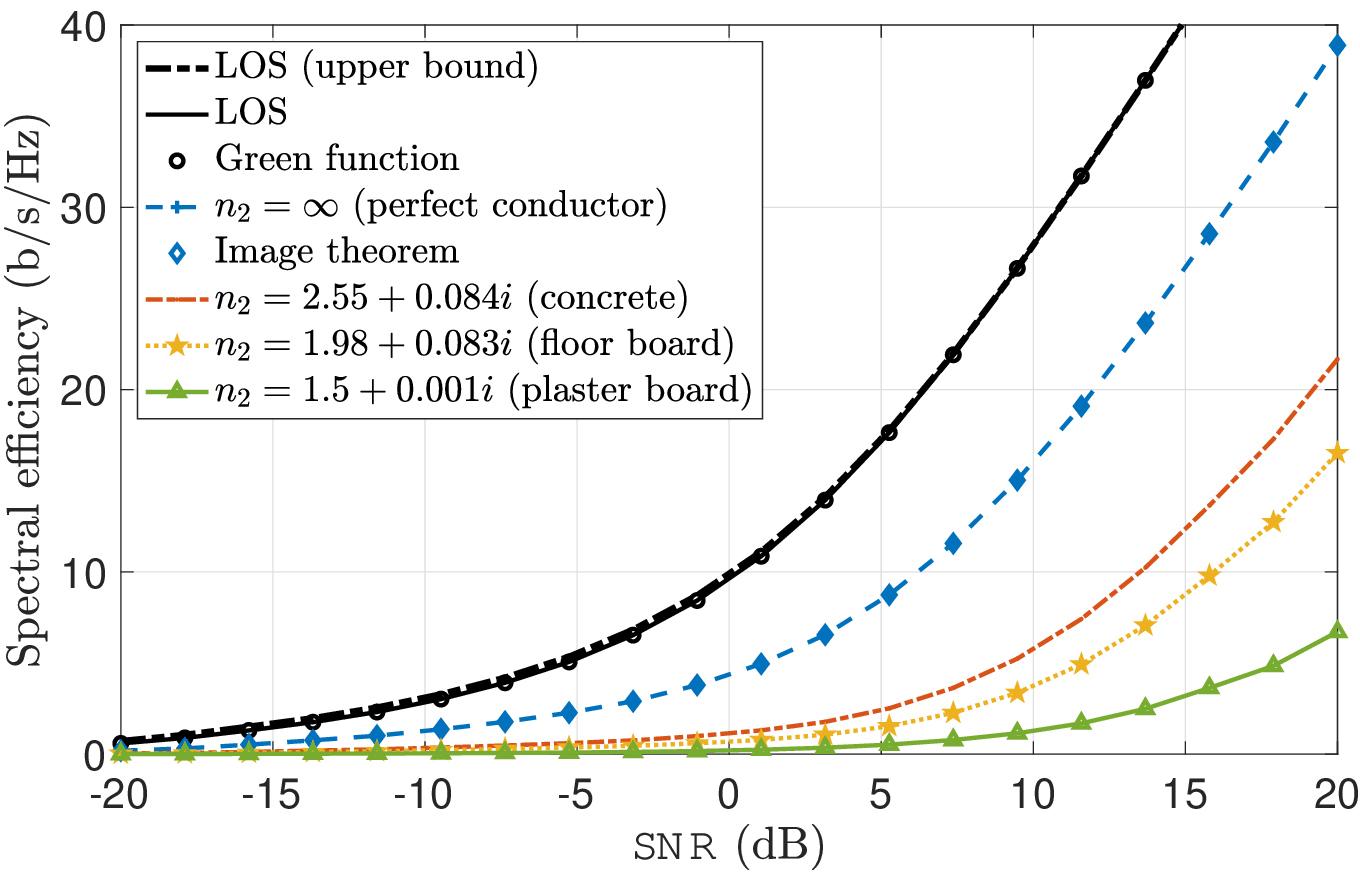} \vspace{-0.0cm}
\caption{Spectral efficiency as a function of SNR for different materials. Non-parallel ULAs with spacing ${\sf d}(D,{\sf SNR},\vartheta)$ for the LOS channel and ${\sf d}(D_{\text e},{\sf SNR},\vartheta_{\text e})$ for the reflected channel.
}\vspace{-0.0cm}
\label{fig:Fig13}
\end{figure}

Second, we turn to spatial selectivity. Consider oblique incidence on a vertical plane, not necessarily the $xz$-plane. Its projected views on the $yz$-plane and on the $xz$-plane are illustrated in Figs.~\ref{fig:Fig11_top} and~\ref{fig:Fig11_side}.
For the side view in Fig.~\ref{fig:Fig11_side}, we define 
\begin{align}
\widehat{D} & = D / \sqrt{1+ \left(\frac{r_{0y}}{r_{0z}}\right)^{\!2}} \\
\widehat{D}_{\text e} & = D_{\text e} / \sqrt{1+ \left(\frac{r_{0y}}{2D_0 -r_{0z}}\right)^{\!2}} ,
\end{align}
which are obtained by projecting their counterparts $D$ and $D_{\text e}$ onto the $xz$-plane; see Fig.~\ref{fig:Fig11_top}.
As sketched in Fig.~\ref{fig:Fig11_model}, shifting the receiver along the $x$-axis is equivalent to rotating the transmitting and receiving ULAs with respect to the $x$-axis by an angle 
\begin{equation} \label{vartheta}
\vartheta = \arccos \! \left( r_{0z}/\widehat{D} \right)
\end{equation}
for the LOS channel, and by another angle
\begin{equation} \label{vartheta_eq}
\vartheta_{\text e} = \arccos \! \left( \frac{2D_0 - r_{0z}}{ \widehat{D}_{\text e}} \right)
\end{equation}
for the reflected channel.
Unlike the power loss, spatial selectivity can be corrected by tailoring the ULA spacing opportunely \cite{Heedong2021}.
To this end, for the LOS channel,
\begin{equation} \label{spacing_theta}
{\sf d}(D,{\sf SNR},\vartheta) = \frac{{\sf d}(D,{\sf SNR})}{\cos(\vartheta)},
\end{equation} 
with ${\sf d}(D,{\sf SNR})$ the optimal antenna spacing for parallel ULAs in \eqref{spacing} whereas, for the reflected channel, ${\sf d}(D_{\text e},{\sf SNR},\vartheta_{\text e})$ with $D_{\text e}$ in \eqref{equivalent_dist_theta} and $\vartheta_{\text e}$ in \eqref{vartheta_eq}.
Compared to parallel ULAs, non-parallel ULAs have antennas that are spaced further apart due to the division by $\cos(\cdot)$ in \eqref{spacing_theta}. The potential DOF thereby shrink for ULAs tilted sideways. 
At high SNR, since $\eta({\sf SNR})=1$ in \eqref{spacing}, ${\sf d}(D,{\sf SNR})$ reduces to ${\sf d}(D)$ in \eqref{Rayleigh} leading to full-rank channel matrices for the LOS and reflected channels.
This property is validated in Fig.~\ref{fig:Fig12} for an $x$-oriented ULA located at $\vect{r}_0 = (1,4,10)$~m, i.e., for $\vartheta=5.3^\circ$ and $\vartheta_{\text e}=2.8^\circ$.
Finally, the spectral efficiency with ULA spacings optimized at every SNR for the LOS and reflected transmissions is also shown in Fig.~\ref{fig:Fig13} for different materials.



\section{Implications for Ray Tracing Algorithms} \label{sec:ray_tracing}

\begin{figure}
\centering\vspace{-0.0cm}
\includegraphics[width=.999\linewidth]{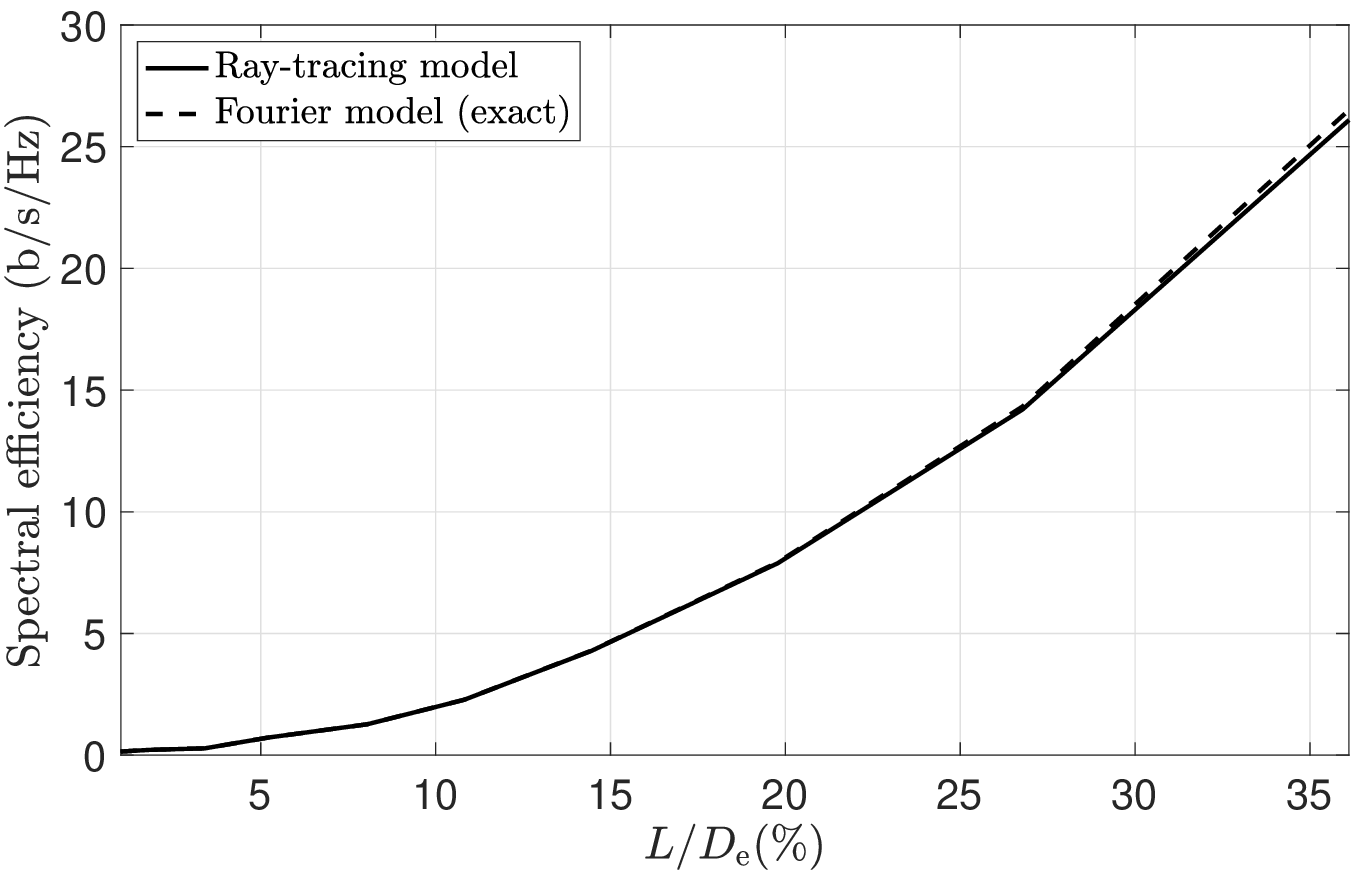} \vspace{-0.0cm}
\caption{Spectral efficiency as a function of the array apertures over equivalent distance ratio at ${\sf SNR} = 0$~dB when the material is concrete.
The antenna spacing is optimized for the reflected transmission.
}\vspace{-0.0cm}
\label{fig:Fig14}
\end{figure}



NLOS connectivity is typically established via multiple reflections involving possibly distinct materials and orientations. Analysis becomes unwieldy in such general settings and the recourse are numerical algorithm such as ray tracing \cite{MolischBook}.
Our setup provides insights into the mechanisms involved at each stage of reflection. 

Our exact channel model describes the reflected propagation as an LSI filtering, whereas the ray-tracing model regards the convolving response as an impulse weighted by the reflectivity coefficient in~\eqref{R_theta} at $\theta_{\text i} = \theta_0$.
To appreciate the difference between the exact and the approximated method (ray-tracing) one should increase the array apertures $L_{\text{r}}$ and $L_{\text{t}}$ for a given communication range $D_{\text{e}}$, thus violating the sufficient condition for the reflected transmission to be paraxial. To this end, Fig.~\ref{fig:Fig14} depicts the spectral efficiency of the reflected channel between two ULAs of apertures $L_{\text{r}}=L_{\text{t}}=L$ as a function of $L/D_{\text{e}}$ at ${\sf SNR} = 0$~dB.  
The ULAs are separated by ${D = 2}$~m and the surface distance is ${D_0 = 3}$~m so that $D_{\text{e}} = 4$~m. In turn, the antenna spacing is optimized for the reflected transmission, which implies array apertures linearly increase with the number of antennas.

The ray-tracing curve yields a tight match with the exact one, except for the regime where the two arrays have an aperture $L$ comparable to the range $D_{\text{e}}$ of the reflected transmission.
Hence, ray tracing algorithms leveraging the paraxial approximation offer a good fit to reality, as also supported by the robustness of the underlying approximation against changes in the propagation geometry.

\section{Conclusion} \label{sec:conclusions}

      \begin{figure*}[th!]
\begin{equation} \tag{47} \label{impulse_response_cylindrical}
h(\delta_\rho;r_z,s_z) = 
\begin{cases} 
\begin{aligned} \displaystyle
& \frac{\kappa_1 \eta_1}{4\pi} \int_{0}^{\kappa_1} \kappa_{\rho} \frac{J_0(\kappa_\rho \delta_\rho)}{\kappa_{1z}} \left(  e^{-\imagunit \kappa_{1z} (r_z-s_z)} +  R(\kappa_{\rho}) e^{-\imagunit \kappa_{1z} (r_z+s_z-2D_0)} \right)  d\kappa_\rho   & r_z < -R_0 \\ \displaystyle
& \frac{\kappa_1 \eta_1}{4\pi} \int_{0}^{\kappa_1}  \kappa_{\rho} \frac{J_0(\kappa_{\rho} \delta_\rho)}{\kappa_{1z}} \left(e^{\imagunit \kappa_{1z} (r_z-s_z)} +  R(\kappa_{\rho}) e^{- \imagunit \kappa_{1z} (r_z+s_z-2D_0)} \right) d\kappa_\rho  & R_0 < r_z \le D_0 
\end{aligned}
\end{cases} 
\end{equation}
\hrule
\end{figure*}

Through a physics-based formulation, 
we have confirmed that reflection off a large and smooth planar surface, say a wall or ceiling, can serve as alternatives to LOS for wide-aperture MIMO communication.
 With respect to an LOS link, a reflected counterpart exhibits:
\begin{itemize}
\item
A power loss determined by the additional range and by the share of incident power not reflected by the surface.
\item
A reduction in the number of DOF because of the antenna spacing tailored to the LOS link being smaller than the one that the reflected link would require at the same SNR.
\end{itemize}

If the arrays are outright configured for the reflected transmission, then the second effect is corrected. 
The above observations bode well for flexible LOS MIMO communication aided by reflections, with further work required to determine the impact of surface finiteness and roughness. 
This paper ignores mutual coupling effects among antenna elements, which are most impactful at sub-wavelength spacings \cite{Nossek}. This ought not to be the case for wide-aperture MIMO that envisions electrically large antenna spacings, with follow-up studies needed to confirm this hypothesis.



Connection with the image theorem that underlies ray tracing showed that, with non-planar wavefronts, the image of the transmitter is blurred by the convolution with a response modeling the 
not perfect reflectivity of the surface.
Ray tracing ignores this blurring, which is to say it regards the convolving response as an impulse. However, our findings show that only for very large arrays
does the response depart from an impulse, justifying the use of ray tracing algorithms \cite{Rangan2022} in most situation.

\appendix
\section*{Generation of the MIMO channel matrix} \label{app:numerical_generation}
        
For the sake of compactness, let us define the space-lag variable ${\vect{\delta} = \vect{r}-\vect{s}}$ with coordinates ${\delta_x = r_x-s_x}$ and ${\delta_y = r_y-s_y}$, indicating the displacement between source and receiver on the $z$-plane.
Due to circular symmetry of $H(\vect{\kappa};r_z,s_z)$ in \eqref{wavenumber_response}, we eliminate the azimuthal dependance of the channel impulse response by evaluating \eqref{impulse_response_Fourier} at $(\delta_\rho,0)$, $\delta_\rho = \|\vect{\delta}\|$. The result is reported in \eqref{impulse_response_cylindrical} for any given $s_z$ and $r_z$ where we introduced $\kappa_\rho = \|\vect{\kappa}\| \in[0,\kappa_1]$ within $\vect{\kappa}\in\mathcal{D}$.
Hence, the impulse response is invariant under any affine transformation that preserves the distance between source and receiver on the $z$-plane. Eq. \eqref{impulse_response_cylindrical} is a Sommerfeld-type integral \cite[Eq.~2.2.30]{ChewBook}. This describes the received field as integral superpositions of cylindrical waves times an upgoing or downgoing plane wave in the $z$-direction. 
Analytical solutions of \eqref{impulse_response_cylindrical} are hardly available and problem-specific \cite[Ch.~2.7.3]{ChewBook}. Hence, we resort to a numerical integration procedure that accounts for the singularities on the complex $\kappa_{\rho}$-plane.\footnote{The error of a numerical integration routine
is proportional to the derivatives of the integrand and are unbounded near a singularity \cite{HildebrandBook}.}

Assuming the analyticity of the integrand, we can invoke Cauchy's integral theorem and deform the contour integration path to avoid singularities. The integral value is unchanged along this new integration path. This should lie in the fourth orthant due to ${\Re(\kappa_\rho) \ge 0}$ and Sommerfeld radiation conditions (i.e., ${\Im(\kappa_{1z}) \ge 0}$ and ${\Re(\kappa_{1z}) \ge 0}$) that ensures convergence of the improper integral in \eqref{impulse_response_cylindrical} \cite[Ch.~2.2.3]{ChewBook}.\footnote{The half-planes ${\Im(\kappa_{1z}) = 0}$ and ${\Re(\kappa_{1z}) = 0}$ map to the hyperbola \cite[Eq.~2.2.33]{ChewBook} in the complex $\kappa_\rho$ plane; see~\cite[Fig.~2.2.8]{ChewBook}.}
We follow \cite{Paulus} and choose a semi-elliptical integration path $\mathcal{C}$ that goes around the pole singularities with semi-axes of the ellipse 
 chosen as \cite{Paulus}
\setcounter{equation}{47} 
\begin{align} \label{semi_ellipse}
\kappa_\rho^{\text{maj}} = (\kappa_1 + \kappa_2)/2  & \qquad
\kappa_\rho^{\text{min}} = \kappa_\rho^{\text{maj}}/10^3
\end{align}
so that the contour of $\mathcal{C}$ is sufficiently away from the singularity but $\kappa_\rho$ is small enough for the argument of the Bessel function in \eqref{impulse_response_cylindrical} to ensure controlled oscillations.
For complex integration, we parametrize the curve as $\kappa_{\rho}(\theta) : [\pi, 2\pi) \to \mathcal{C}$ where $\kappa_{\rho}(\theta) = \kappa_{\rho}^\prime(\theta) + \imagunit \kappa_{\rho}^{\dprime}(\theta)$ with 
\begin{align}
\kappa_{\rho}^\prime(\theta) = \frac{\kappa_\rho^{\text{maj}}}{2} (1 + \cos(\theta)) & \;\;\;  \kappa_{\rho}^{\dprime}(\theta) = \frac{\kappa_\rho^{\text{min}}}{2} \sin(\theta),
\end{align}
leading to \cite[Ch.~10.5]{HildebrandBook}
\begin{equation} \label{contour}
\int_{\mathcal{C}} f(\kappa_{\rho}) \, d\kappa_{\rho} =  \int_{0}^\pi f(\kappa_{\rho}(\theta)) \left|\frac{\partial\kappa_{\rho}(\theta)}{\partial \theta}\right|  \, d\theta
\end{equation}
where $f(\kappa_{\rho})$ is the integrand of \eqref{impulse_response_cylindrical} and the Jacobian is
\begin{equation}
\frac{\partial\kappa_{\rho}(\theta)}{\partial \theta} = \frac{1}{2} \left(-\kappa_\rho^{\text{maj}} \sin(\theta) + \imagunit \kappa_\rho^{\text{min}} \cos(\theta) \right).
\end{equation}

The presented numerical generation procedure performs superbly as long as the transverse distance $\rho$ is not too large compared to the wavelength $\lambda$. Numerical simulations show no issue for $\rho < 18$~m at $60$~GHz, i.e., $\delta_\rho/\lambda < 3600$. For larger $\delta_\rho$, the integrand in \eqref{impulse_response_cylindrical} becomes a rapidly oscillating function of $\kappa_\rho$, due to the large variations into the Bessel function, and $\mathcal{C}$ must be chosen according to the steepest descent path \cite[Ch.~2.7.3]{ChewBook}.

\bibliographystyle{IEEEbib}
\bibliography{refs}

\begin{thebibliography}{10}

\bibitem{PizzoVTC22}
A.~Pizzo, A.~Lozano, S.~Rangan, and T.~L. Marzetta,
\newblock ``Line-of-sight {MIMO} via reflection from a smooth surface,''
\newblock in {\em IEEE Veh. Technol. Conf. (VTC)}, 2022.

\bibitem{Sundeep2014}
S.~Rangan, T.~S. Rappaport, and E.~Erkip,
\newblock ``Millimeter-wave cellular wireless networks: Potentials and
  challenges,''
\newblock {\em Proc. IEEE}, vol. 102, no. 3, pp. 366--385, 2014.

\bibitem{Rappaport2015}
T.~S. Rappaport, G.~R. MacCartney, M.~K. Samimi, and S.~Sun,
\newblock ``Wideband millimeter-wave propagation measurements and channel
  models for future wireless communication system design,''
\newblock {\em IEEE Trans. Commun.}, vol. 63, no. 9, pp. 3029--3056, 2015.

\bibitem{Rappaport2019}
T.~S. Rappaport, Y.~Xing, O.~Kanhere, S.~Ju, A.~Madanayake, S.~Mandal,
  A.~Alkhateeb, and G.~C. Trichopoulos,
\newblock ``Wireless communications and applications above 100 {GHz}:
  Opportunities and challenges for {6G} and beyond,''
\newblock {\em IEEE Access}, vol. 7, pp. 78729--78757, 2019.

\bibitem{Rodwell2011}
E.~Torkildson, U.~Madhow, and M.~Rodwell,
\newblock ``Indoor millimeter wave {MIMO}: Feasibility and performance,''
\newblock {\em IEEE Trans. Wireless Commun.}, vol. 10, no. 12, pp. 4150--4160,
  2011.

\bibitem{Bohagen}
Frode Bohagen, Pal Orten, and Geir~E. Oien,
\newblock ``On spherical vs. plane wave modeling of line-of-sight {MIMO}
  channels,''
\newblock {\em IEEE Trans. Commun.}, vol. 57, no. 3, pp. 841--849, 2009.

\bibitem{9422343}
H.~Do, S.~Cho, J.~Park, H.-J. Song, N.~Lee, and A.~Lozano,
\newblock ``Terahertz line-of-sight {MIMO} communication: Theory and practical
  challenges,''
\newblock {\em IEEE Commun. Magazine}, vol. 59, no. 3, pp. 104--109, 2021.

\bibitem{Rangan2022}
Yaqi Hu, Mingsheng Yin, Sundeep Rangan, and Marco Mezzavilla,
\newblock ``Parametrization of high-rank line-of-sight {MIMO} channels with
  reflected paths,''
\newblock in {\em 2022 IEEE 23rd Int. Workshop Signal Process. Advances
  Wireless Commun. (SPAWC)}, 2022, pp. 1--5.

\bibitem{OpticsBook}
M.~Born and E.~Wolf,
\newblock {\em Principles of Optics},
\newblock Pergamon Press, 6 edition, 1980.

\bibitem{ChewBook}
W.~C. Chew,
\newblock {\em Waves and Fields in Inhomogenous Media},
\newblock Wiley-IEEE Press, 1995.

\bibitem{PlaneWaveBook}
T.~B. Hansen and A.~D. Yaghjian,
\newblock {\em Plane-Wave Theory of Time-Domain Fields},
\newblock Wiley-IEEE Press, New York, 1999.

\bibitem{PizzoTSP21}
A.~Pizzo, A~de~J. Torres, L.~Sanguinetti, and T.~L. Marzetta,
\newblock ``Nyquist sampling and degrees of freedom of electromagnetic
  fields,''
\newblock {\em IEEE Trans. on Signal Process.}, pp. 1--12, 2022,
  doi={10.1109/TSP.2022.3183445}.

\bibitem{PizzoJSAC20}
A.~{Pizzo}, T.~L. {Marzetta}, and L.~{Sanguinetti},
\newblock ``Spatially-stationary model for {Holographic} {MIMO} small-scale
  fading,''
\newblock {\em IEEE J. Sel. Areas Commun.}, vol. 38, no. 9, pp. 1964--1979,
  2020.

\bibitem{PizzoIT21}
A.~Pizzo, L.~Sanguinetti, and T.~L. Marzetta,
\newblock ``Spatial characterization of electromagnetic random channels,''
\newblock {\em IEEE Open J. Commun. Soc.}, vol. 3, pp. 847--866, 2022.

\bibitem{HeedongISIT}
H.~Do, N.~Lee, and A.~Lozano,
\newblock ``Capacity of line-of-sight {MIMO} channels,''
\newblock in {\em IEEE Int. Symp. Inf. Theory (ISIT)}, 2020, pp. 2044--2048.

\bibitem{Heedong2021}
H.~Do, N.~Lee, and A.~Lozano,
\newblock ``Reconfigurable {ULAs} for line-of-sight {MIMO} transmission,''
\newblock {\em IEEE Trans. Wireless Commun.}, vol. 20, no. 5, pp. 2933--2947,
  2021.

\bibitem{MolischBook}
A.~F. Molisch,
\newblock {\em {Wireless Communications}},
\newblock Wiley-IEEE Press, 2 edition, 2011.

\bibitem{OrfanidisBook}
S.~J. Orfanidis,
\newblock {\em Electromagnetic Waves and Antennas},
\newblock 2016. [Online] Available:
  \url{https://www.ece.rutgers.edu/~orfanidi/ewa/}.

\bibitem{Sato}
K.~Sato, H.~Kozima, H.~Masuzawa, T.~Manabe, T.~Ihara, Y.~Kasashima, and
  K.~Yamaki,
\newblock ``Measurements of reflection characteristics and refractive indices
  of interior construction materials in millimeter-wave bands,''
\newblock in {\em IEEE Veh. Technol. Conf. (VTC)}, 1995.

\bibitem{BalanisBook}
C.~A Balanis,
\newblock {\em Antenna Theory: Analysis and Design},
\newblock Wiley-Interscience, 4 edition, 2005.

\bibitem{tulino2004mimo}
A~Tulino, A.~Lozano, and S.~Verdu,
\newblock ``{MIMO} capacity with channel state information at the
  transmitter,''
\newblock in {\em IEEE Int'l Symp. Spread Spectrum Techniques and Applications
  (ISSSTA)}, 2004.

\bibitem{LozanoBook}
R.~W. Heath~Jr. and A.~Lozano,
\newblock {\em Foundations of {MIMO} Communication},
\newblock Cambridge University Press, 2018.

\bibitem{Nossek}
Michel~T. Ivrlač and Josef~A. Nossek,
\newblock ``Toward a circuit theory of communication,''
\newblock {\em IEEE Trans. Circuits Syst.}, vol. 57, no. 7, pp. 1663--1683,
  2010.

\bibitem{HildebrandBook}
F.~B. Hildebrand,
\newblock {\em {Advanced Calculus for Applications}},
\newblock Prentice Hall, 1962.

\bibitem{Paulus}
M.~Paulus, P.~Gay-Balmaz, and O.~J.~F. Martin,
\newblock ``Accurate and efficient computation of the {Green}'s tensor for
  stratified media,''
\newblock {\em Phys. Rev. E}, vol. 62, pp. 5797--5807, Oct 2000.

\end{thebibliography}

\end{document}